\documentclass[%
 reprint,
 superscriptaddress,
 amsmath,amssymb,aps,prl
]{revtex4-2}

\usepackage{graphicx}
\usepackage{dcolumn}
\usepackage{bm}

\usepackage[draft]{changes}
\usepackage{todonotes}

\setuptodonotes{inline,size=\tiny,color=orange}

\definechangesauthor[color=red]{EW}

\begin{document}


\title{Fundamental and second-subharmonic Autler-Townes splitting in classical systems}
\author{Ahmed A. Barakat}
\email{Contact author: ahmed.barakat@tum.de}
 \affiliation{%
 School of Computation, Information and Technology, Technical University of Munich, Munich, Germany.}
 \affiliation{Design and Production Engineering Dept., Faculty of Engineering, Ain Shams University, Cairo, Egypt.}
 \affiliation{Munich Center for Quantum Science and Technology (MCQST), Munich, Germany}
\affiliation{TUM Center for Quantum Engineering (ZQE), Garching, Germany}



\author{Avishek Chowdhury}
\affiliation{%
 School of Computation, Information and Technology, Technical University of Munich, Munich, Germany.}
 
 \author{Anh Tuan Le}%
 \affiliation{%
 School of Computation, Information and Technology, Technical University of Munich, Munich, Germany.}
\author{Eva M. Weig}
\affiliation{%
 School of Computation, Information and Technology, Technical University of Munich, Munich, Germany.}
 \affiliation{Munich Center for Quantum Science and Technology (MCQST), Munich, Germany}
\affiliation{TUM Center for Quantum Engineering (ZQE), Garching, Germany}


\date{\today}

\begin{abstract}
The dynamic Stark effect and the Autler–Townes splitting (ATS) are hallmarks of driven two-level systems. We establish a direct correspondence between these quantum phenomena and the parametric normal mode splitting in coupled classical oscillators. This gives rise to a second-subharmonic ATS under a two-tone parametric drive. We find excellent agreement between the theory and the vibrations of a nanomechanical two-mode system, capturing both the fundamental and second-subharmonic ATS, and allowing quantitative extraction of the modal coupling irrespective of the degree of modal hybridization.
\end{abstract}

\maketitle

\textit{Introduction - }The dynamic (ac) Stark effect, first identified by Autler and Townes \cite{Autler1955}, describes the field-induced modification of energy levels in coherently driven quantum systems. In atomic and optical physics, this light shift forms the basis of optical lattices and lattice clocks \cite{Bloch2008,Ludlow2015}, and governs coherent optical control in semiconductors \cite{Sie2015}. Controlled Stark shifts are likewise essential for qubit manipulation and field sensing in Rydberg and superconducting architectures \cite{Krantz2019,Saffman2010}, for modulation of excitons \cite{Sie2015,Jundt2008,Unold2004}, and for inducing dressed states in cavity and circuit quantum electrodynamics and cavity optomechanics \cite{Blais2021,Aspelmeyer2014}. Although originally described semiclassically, its quantized description arises naturally within the Jaynes–Cummings model~\cite{Gerry_Knight_2023}.

The Autler-Townes Splitting (ATS) is an immediate consequence of the dynamic Stark effect when the field is tuned at the atom transition frequency~\cite{Autler1955}. An analogous phenomenology is found in classical systems, where a parametrically induced modal coupling produces the so-called parametric normal mode splitting (PNMS)~\cite{Frimmer_PNMS_ATS}. This is a consequence of the similarity of the dynamical structure due to the underlying time-varying potential; however, a rigorous mathematical mapping beyond the rotating-wave approximation is yet to be established. 

In the classical regime, two linearly (but not necessarily strongly) coupled mechanical modes produce PNMS, upon the additional application of a parametric drive~\cite{okamoto_coherent_2013}. The approximate relation for PNMS in Ref.~\cite{okamoto_coherent_2013} has been applied to achieve tunable and controllable coupling in various nano- and micromechanical systems~\cite{li_parametric_2016,zhu_coherent_2017,mathew_dynamical_2016,liu_optical_2015,prasad_gate_2019,okamoto_coherent_2013,okamoto_strongly_2016,huang_nonreciprocal_2016,halg_strong_2022}. However, the relation between the PNMS splitting width $2g$ and the modal coupling strength $\lambda$ has not been appreciated to date. In the case of dispersively coupled modes the parametric nature of the coupling induces the same phenomenon~\cite{dobrindt_parametric_2008,wilson-rae_cavity-assisted_2008}. This is, e.g., apparent in strongly coupled cavity optomechanical systems~\cite{aspelmeyer_cavity_2014, groblacher_observation_2009}. Note that the term avoided crossing is often used in the literature interchangeably with normal mode splitting~\cite{Guo2023,groblacher_observation_2009}. Here, we will adhere to the following convention: The term "avoided crossing" will be solely employed for the case of lifting the degeneracy of the natural eigenfrequencies when tuned on resonance via an external parameter. This phenomenon is also known in mechanics as "mode-veering"~\cite{leissa_curve_1974,pierre_mode_1988}. The term "parametric normal mode splitting" will refer to the splitting of \emph{each} of the two linearly coupled normal modes into two branches under the application of an additional parametric drive tuned to their frequency difference.

In this work, we show that the dynamics of a driven two-level system is in one-to-one correspondence with the slow-amplitude dynamics of parametrically coupled classical harmonic oscillators. Both systems exhibit a fundamental splitting at the transition (or difference) frequency, in addition to higher-order splittings at its odd fractions~\cite{Autler1955}. 
Moreover, we show that applying a two-tone parametric drive with a frequency ratio of $1:2$ introduces an additional splitting at half the modal difference frequency - an effect that has been observed before in parametrically driven classical systems~\cite{mathew_dynamical_2016,okamoto_coherent_2013}, but was not considered in the original ATS framework~\cite{Autler1955}. This \emph{second-subharmonic ATS} can be linked to signatures of bichromatically driven atoms~\cite{Rudolph:98}.

\begin{figure}[!t]
	\centering
	\includegraphics[width=\linewidth]{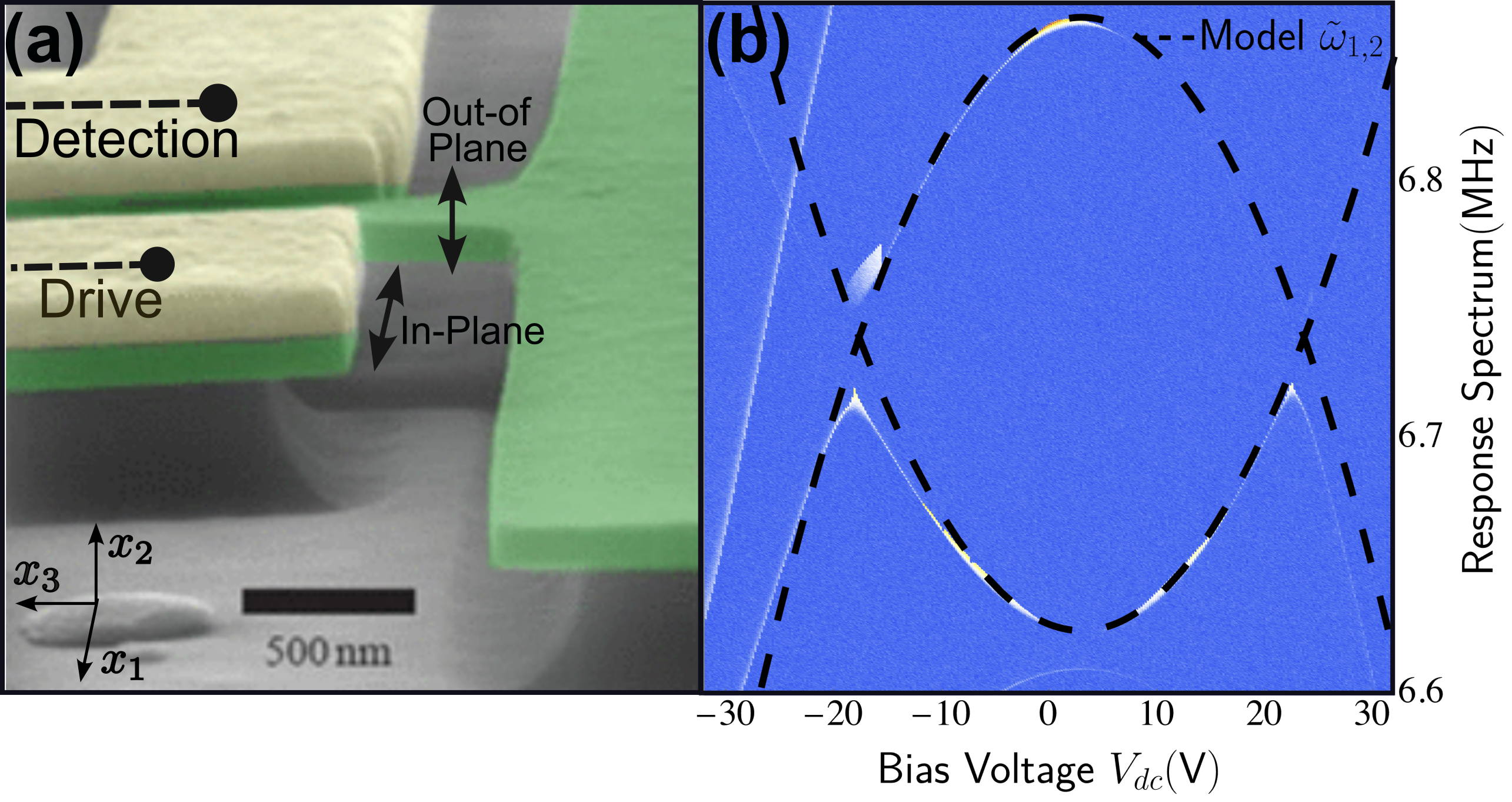}	
	\caption{(a) Scanning electron micrograph of a SiN nanostring (green) flanked by two golden electrodes (yellow), each connected either to the drive and the detection circuitry. (b) Spectrogram of the IP and OOP frequencies of the nanostring as a function of the bias voltage $V_{dc}$ varied from –32 V to +32 V. The data are overlaid with a model of the uncoupled, tunable eigenfrequencies $\tilde{\omega}_i$ (dashed black lines), revealing two avoided crossing points at $V_{dc}\simeq \pm20$ V.}
	\label{fig:efeld}
\end{figure}

In addition, we present a thorough theoretical model to describe the PNMS from a basic classical model using the multiple scales perturbation method~\cite{Nayfeh2000,ramakrishnan_primary_2022}, relating it to the coupling strength and the associated Bloch-Siegert effect~\cite{cohen2024atom}. This facilitates the quantification of the coupling strength even in the weak limit, where the avoided crossing effect is no longer resolved~\cite{Rodriguez2016}. The results are experimentally validated by exploring the two fundamental flexural eigenmodes of a strongly pre-stressed nanomechanical string resonator, denoted as out-of-plane (OOP) and in-plane (IP) modes, which are dielectrically coupled by a dc voltage applied between two adjacent electrodes (see Fig.~\ref{fig:efeld}(a))~\cite{Faust2012}. The modes of the string are driven by a white-noise force, while the parametric coupling is induced by modulating both modes at their difference frequency with an additional tone.

\textit{Theory and modeling - }
The equations of motion of the IP ($x_1$) and OOP ($x_2$) fundamental flexural eigenmode of the nanostring surrounded by the inhomogeneous electric field between the electrodes (see Fig.~\ref{fig:efeld}(a)) are
\begin{equation}\label{eq:electromech_red}
	\begin{aligned}
		\ddot x_1 + 2\beta_1 \dot x_1+ (\tilde{\omega}_1^2 +\eta_1(t))x_1+(\lambda^2+\eta_c(t))x_2=&f(t),\\
		\ddot x_2 + 2\beta_2 \dot x_2+ (\tilde{\omega}_2^2 +\eta_2(t))x_2+ (\lambda^2+\eta_c(t))x_1=&f(t).
	\end{aligned}
\end{equation}
with $\tilde{\omega}_{1,2}$, $\lambda$ representing the dielectrically tunable eigenfrequencies and coupling strength, respectively, and $\beta_{1,2}$ are the amplitude damping rates ~\cite{unterreithmeier_universal_2009,rieger_frequency_2012,faust_nonadiabatic_2012}. A driving force $f(t)$, as well as inherently induced parametric actuation terms $\eta_{1,2}(t),\eta_c(t)$ are applied. 
Under weak driving, nonlinearities and higher-order coupling terms~\cite{Bachtold2022} do not influence the system dynamics and are omitted for the sake of simplicity. As a result of the dielectric coupling, $x_1$ and $x_2$ hybridize, and the observable modes are found after diagonalization of Eq.~(\ref{eq:electromech_red}), yielding the equations of motion in the hybridized coordinates $q_i$ ($i=1,2$)
\begin{equation}\label{eq:electromech_assym}
	\begin{aligned}
		\ddot q_1 + 2\gamma \dot q_1+ (\Omega_1^2+\Sigma_1(t)) q_1+\zeta\Lambda(t) q_2=&F(t),\\
		\ddot q_2 + 2\gamma \dot q_2+ (\Omega_2^2+\Sigma_2(t)) q_2+\>\>\Lambda(t) q_1=&F(t),
	\end{aligned}
\end{equation}
with damping $\gamma$, modal eigenfrequencies $\Omega_{i}$, drive $F(t)$, parametric actuation $\Sigma_{i}(t)$ and time-dependent parametric coupling $\Lambda(t)$ oscillating at frequency $\Omega_p$, and a scaling factor $\zeta$~\cite{supp}. The relation between the modal eigenfrequencies of the hybridized modes $\Omega_i$ and the original modes is
\begin{equation}\label{eq:eigfreq_hybrid}
	\Omega^2_{1,2}=\frac{1}{2}(\tilde{\omega}_1^2+\tilde{\omega}_2^2)\mp\frac{1}{2} \sqrt{(\tilde{\omega}_2^2-\tilde{\omega}_1^2)^2 + 4\lambda^4}.
\end{equation}

Similarly, the coordinate transformation establishes a relation between the amplitude of the parametric coupling $\Lambda$ and the original coupling strength $\lambda$~\cite{supp}

\begin{equation}\label{eq:Lambda_lambda_rel}
	\Lambda_\zeta=\frac{\sqrt{\zeta}V_{ac}(t)\delta\omega^2}{V_{dc}}\left(\frac{\delta\omega^2+(c_{11}-c_{22})V_{dc}^2}{\sqrt{4\lambda^4+((c_{11}-c_{22})V_{dc}^2+\delta\omega^2)^2}}-1\right),
\end{equation}
where $\Lambda_\zeta=\sqrt{\zeta}\Lambda$, $\delta\omega^2=(\tilde{\omega}_2^2-\tilde{\omega}_1^2)|_{V_{dc}\approx 0}$, and $c_{ii}$ are the tuning coefficients for $\tilde{\omega}_{i}$ with $V_{dc}$ (see Eq.~(S23) of~\cite{supp}).

At given $\Omega_{i}(V_{dc})$, the multiple-scale method is applied up to the second order for $\Omega_p\simeq \delta \Omega=|\Omega_2-\Omega_1|\ll \Omega_1,\Omega_2$, as discussed in~\cite{supp}. The splitting width is deduced to be
\begin{equation}\label{eq:2g-Lambda-BSeffect}
    2g=\tilde{\Lambda}\sqrt{1+\Bigg(\frac{\tilde{\Lambda}}{4\>\delta\Omega}\Bigg)^2},
\end{equation}
where $\tilde{\Lambda}=\Lambda_\zeta/(2\sqrt{\Omega_1\Omega_2})$ is the Rabi frequency. The last equation reflects the ATS expression~\cite{Autler1955}, where the second term under the root is found to represent the Bloch-Siegert shift~\cite{supp,cohen2024atom,Yan_BSEffect}. This resemblance emerges in fact from a correspondence between the dynamical structures that induce both ATS and PNMS. A canonical change of variables $(q_i,\dot q_i)\mapsto (C_i,C_i^*)$ under a slow-envelope approximation, converts Eq.~(\ref{eq:electromech_assym}) into the semiclassical ATS form~\cite{Autler1955}
\begin{equation}
    \begin{aligned}
        i\dot{C}_1(t)&=\Omega_1 C_1(t)+\tilde{\Lambda}\cos(\Omega_pt)C_2(t),\\
        i\dot{C}_2(t)&=\Omega_2 C_2(t)+\tilde{\Lambda}\cos(\Omega_pt)C_1(t),
    \end{aligned}
\end{equation}
where $\Sigma_i=0,\gamma=0,\zeta=1$ are assumed for simplicity. The Floquet Hamiltonians~\cite{Shirley_ATS} of both systems yield the same spectra of quasi-energies (c.f. Fig.~S1 of~\cite{supp}), giving rise to ATS (or PNMS) at the frequencies $\Omega_p=\frac{1}{n}\delta\Omega$, and $n=1,3,5,...$ . The case of second-subharmonic ATS, i.e., when $n=2$, is treated separately in a following section.

 \textit{Measurement -} 
 The PNMS model is applied to the two orthogonally polarized fundamental flexural modes, the OOP and IP modes, 
 of a strongly pre-stressed nanomechanical string resonator. The string is made of stoichiometric silicon nitride and has a length, width and thickness of $65\,\mu$m, $250$\,nm and $100$\,nm, respectively. It is placed between two adjacent gold electrodes for dielectric transduction and control, and read out by means of microwave cavity-assisted heterodyne displacement detection using a coaxial $\lambda / 4$ microwave cavity as described elsewhere~\cite{unterreithmeier_universal_2009,Faust2012, rieger_frequency_2012, le_3d_2023,supp}. Throughout this work, a microwave cavity pump power of $13$\,dBm is applied on resonance with the cavity mode for displacement detection.
 
The nanostring is characterized by probing its resonant response to a coherent resonant drive $V_{ac}$ with a constant amplitude and a swept frequency as a function of the applied dc bias voltage $V_{dc}$. Figure~\ref{fig:efeld}(b) shows the resulting eigenfrequency tuning of the OOP and the IP mode. Clearly, the characteristic approximately quadratic tuning behavior leading to a stiffening (softening) of the OOP (IP) mode, respectively, is observed~\cite{rieger_frequency_2012}. The avoided crossings near $V_{dc}\approx \pm 20$\,V are reminiscent of the strong coupling between OOP and IP~\cite{faust_nonadiabatic_2012}. The response of the system changes strongly with the system parameters. While parts of the spectrum remain below the noise level, other parts exhibit a strongly nonlinear response. Notice that the spurious mode on the left side of the figure corresponds to the IP mode of a shorter string on the sample. 
The response of the OOP and IP mode is modeled using Eq.~(\ref{eq:eigfreq_hybrid}) for $\lambda=0$ and fitted to the tuned eigenfrequencies $\tilde\omega_{1,2}$ in the vicinity of $V_{dc}=0$ to extract the tuning coefficients $c_{ii}$ as shown as black dashed lines in Fig.~\ref{fig:efeld}(b). The magnitude of the avoided crossing is indicative of the coupling $\lambda$ when both modes are tuned on resonance as discussed in detail in \cite{supp}.
 
 To observe the PNMS, the string is exposed to a parametric excitation  $V_{\text{pump}}$ along with a white-noise driving force $V_{\text{noise}}$ and the dc bias voltage $V_{dc}$. The voltages, $V_{dc}$ and $V_{ac}=V_{\text{noise}}+V_{\text{pump}}$, are combined with a bias-T and applied between the two electrodes. That said, the parametric excitation results from the inherent dielectric interaction, as any oscillating voltage $V_{ac}$ will both act as a direct drive and modulate the modes' eigenfrequencies, see \cite{supp} for details. The predominant interaction is set by the choice of frequency: Voltages close to $\tilde\omega_i$ drive mode $i$ on resonance, while their parametric contribution represents a second-order effect and is thus negligible. On the other hand, voltages fulfilling a parametric resonance condition $\Omega_p=|\Omega_i\pm \Omega_j|/n$ with positive integer $n$ produce parametric effects. In our case, we focus on the difference combination frequency condition $\Omega_p=|\Omega_2-\Omega_1|$ which yields the PNMS. This situation is also known as the beam-splitter interaction~\cite{aspelmeyer_cavity_2014}.
\begin{figure}[t]
\includegraphics[width=\linewidth]{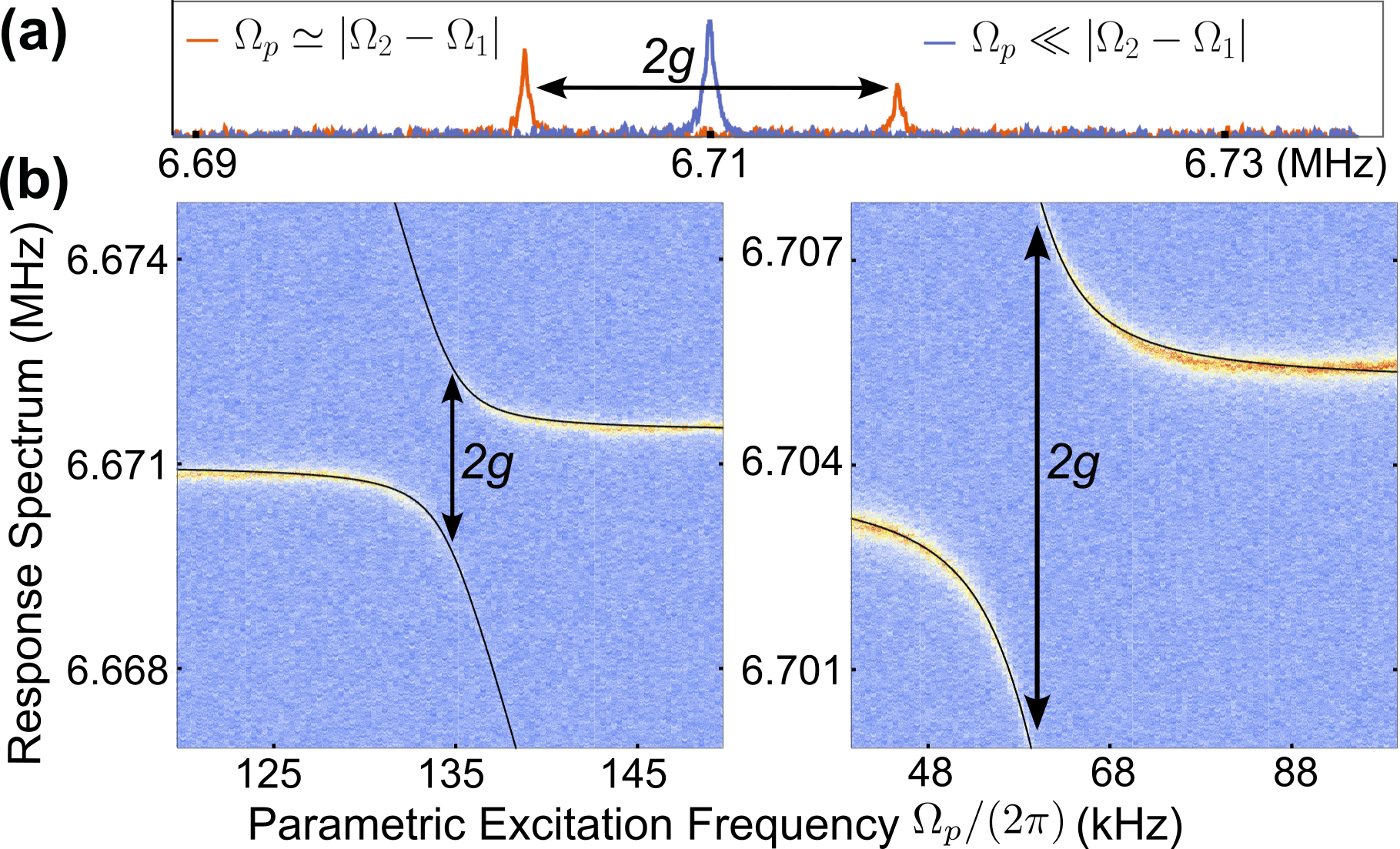}\caption{(a) Unperturbed resonance of $\Omega_1$ at $V_{dc}=-22$V (blue), and split PNMS peaks (orange) under additional parametric perturbation at $\Omega_p\approx\delta\Omega$. (b) Sweeping $\Omega_p$ around $\delta\Omega$ reveals full PNMS branches, at $V_{dc}=-14$\,V (left) and $-20$\,V (right), encoding response amplitudes in color (blue to red). Solid black lines represent the fitted model (see Eq.~(S47)) showing an excellent agreement with the experiment.}
	\label{fig:Exp_PNMS}
\end{figure}

The response of the system near $\Omega_1$ to a parametric modulation with a constant $V_{pump}=2$\,V is depicted in Fig.~\ref{fig:Exp_PNMS} (a) and (b). The blue trace in Fig.~\ref{fig:Exp_PNMS}(a) shows a single peak representing one of the two hybridized modes, $\Omega_1$, excited using the noise drive. The orange trace shows the response when additionally fulfilling the parametric resonance condition, giving rise to PNMS. The apparent normal mode splitting reveals $2g$.

Figure~\ref{fig:Exp_PNMS}(b) shows PNMS measurements for a swept parametric excitation frequency $\Omega_p$ at two different dc bias voltages, representing different coupling strengths and leading to splittings of significantly different widths. Away from the parametric resonance condition, the system converges to $\Omega_1$, while the response branches off in its vicinity. The response of the system is fitted using the PNMS model  (see Eq.~(S47) in ~\cite{supp}). The black solid lines in Fig.~\ref{fig:Exp_PNMS}(b) show excellent agreement with the experimental data. The parametric coupling $\Lambda$ obtained from the fit is converted into $2g$ using Eq.~(\ref{eq:2g-Lambda-BSeffect}). Additionally, the vertical arrows indicate the width of the normal mode splitting $2g$ at the point of minimal separation of the two branches, i.e. $\Omega_p = |\Omega_2-\Omega_1|$, yielding $2$ and $7$\,kHz for Fig.~\ref{fig:Exp_PNMS}(b) (left) and (right), respectively.

\begin{figure}[t]
	\centering
	\includegraphics[width=0.95\linewidth]{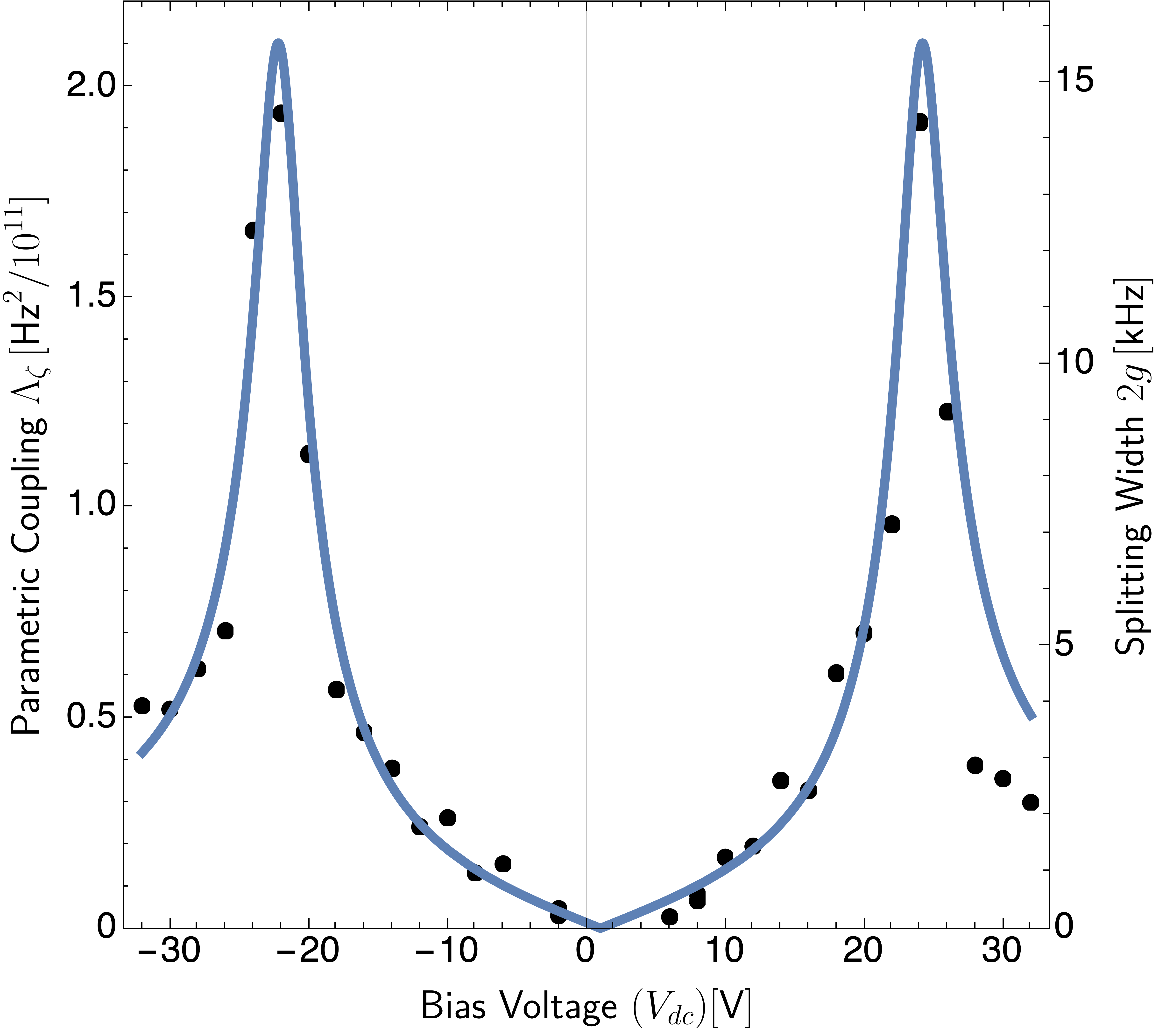}	
	\caption{Splitting width $2g$ and the corresponding scaled parametric coupling $\Lambda_\zeta=\sqrt{\zeta}\Lambda$ values across the bias voltage range (black points), fitted with Eq.~\eqref{eq:Lambda_lambda_rel} based on the ansatz $\lambda= \sqrt{c_\lambda} V_{dc}$ to obtain $c_\lambda$, see Sec.~S8~\cite{supp}.}
	\label{fig:Lambda_theo_exp}
\end{figure}

\textit{Coupling strength estimation - }
The PNMS measurement is repeated for a large number of dc bias voltages between $-32$ to $+32$\,V. The resulting splittings' widths and corresponding parametric coupling values are plotted as a function of $V_{dc}$ as black dots in Fig.~\ref{fig:Lambda_theo_exp}. In this figure, the solid blue line depicts a fit of Eq.~\eqref{eq:Lambda_lambda_rel} to the data, assuming a linear relation~\cite{faust_nonadiabatic_2012}, $\lambda=\sqrt{c_{\lambda}}V_{dc}$, between $\lambda$ and $V_{dc}$~\cite{supp}. We find $c_\lambda \approx 4.41\times10^8 (\text{Hz/V})^2$. Figure~\ref{fig:Lambda_theo_exp} shows a very good agreement between theory and experiment. The axis on the right expresses the result in terms of the width of the normal mode splitting $2g$, taking advantage of $2g\propto\Lambda_\zeta$ according to Eq.~(\ref{eq:2g-Lambda-BSeffect}). Note that $\lambda$ can alternatively be directly calculated via Eq.~\eqref{eq:Lambda_lambda_rel} with Eq.~\eqref{eq:2g-Lambda-BSeffect} highlighting the $\lambda-2g$ relation, see Sec.~S5~\cite{supp}.

Both $2g$ and $\Lambda_\zeta$ exhibit a near-symmetric behavior around $V_{dc}=0$\,V, as expected from the dielectric origin of the coupling. Despite the ever-increasing coupling strength $\lambda$ with $|V_{dc}|$, the splitting width reaches a maximum value at the avoided crossing (c.f. Fig.~\ref{fig:efeld}(b)) before declining further. This can be explained by the hybridization of IP and OOP modes that are coupled by $\lambda$, resulting in Eq.~\eqref{eq:Lambda_lambda_rel}.
It is instructive to compare the magnitude of the coupling strength $\lambda$ obtained from the PNMS with the value of $\lambda$ deduced from the width of the avoided crossing. The avoided crossing produces a splitting $\Gamma/(2\pi) \approx \lambda^2/(2\pi\tilde{\omega}) \approx 47$\,kHz (see Sec.S4~\cite{supp}) at $|V_{dc}|\approx 20$\,V (only the splitting at $-20$\,V is experimentally resolved), c.f. Fig.~\ref{fig:efeld}(b). With $\tilde{\omega}=\tilde{\omega}_1=\tilde{\omega}_2\approx 2\pi\times6.735$\,MHz on resonance, this yields $\lambda/(2\pi)\approx 0.56$ \,MHz.
As detailed above, the PNMS allows for determining $\lambda$ at any bias voltage for known $c_\lambda$. Near resonance, a coupling strength of $\lambda/(2\pi)\approx\sqrt{4.41\times10^8}\times 20=0.42$\,MHz is found in good agreement with the value determined from the avoided crossing.

\begin{figure}[!t]
	\centering
	\includegraphics[width=0.99\linewidth]{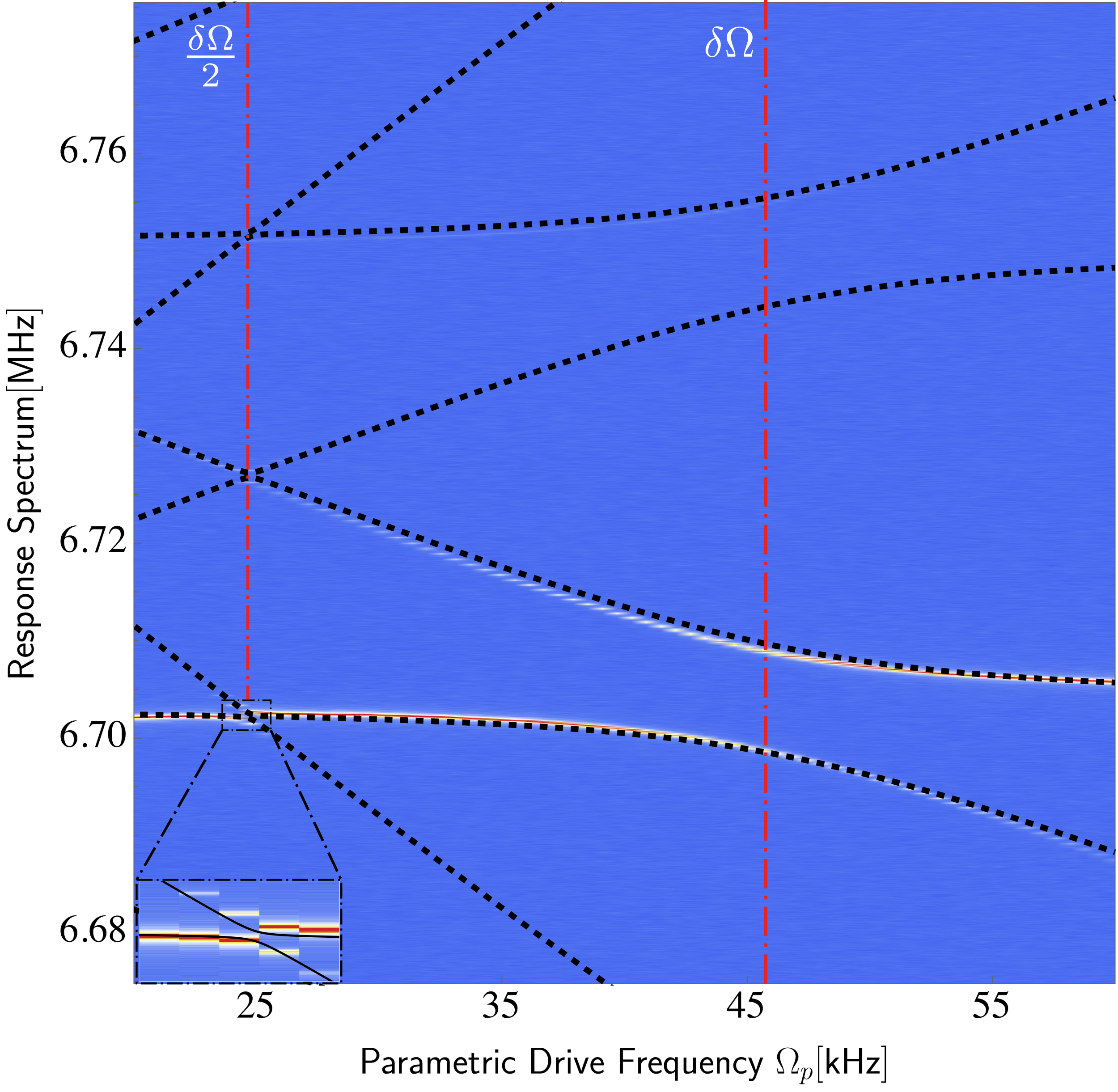}	
	\caption{Experimental map of the quasi-energies at the avoided crossing point $V_{dc}\simeq -20$\,V by sweeping $\Omega_p$, showing the fundamental ATS at $\Omega_p=\delta \Omega$ and the second-subharmonic ATS at $\delta \Omega/2$ (magnified in the inset). The splittings of $\Omega_1$ are clearly resolved, while those of $\Omega_2$ lie mostly under the noise level. The dashed black lines show the Floquet model for a two-tone drive. }
	\label{fig:SecondHarm}
\end{figure}

\textit{Second-subharmonic ATS -}
The ATS-PNMS correspondence, discussed above, suggests that the PNMS experiment should map the dynamic Stark quasi-energies. The dynamic Stark effect of a single-tone $\Lambda(t)$ induces splittings at $\delta\Omega,\delta\Omega/3,\delta\Omega/5,...$~\cite{Autler1955}. In the Floquet representation~\cite{Shirley_ATS}, the single-tone parametric drive couples only adjacent Fourier components, so that the even and odd sectors of the Floquet space decouple, suppressing all even-order resonances~\cite{Mkhitaryan2019}. As depicted in Fig.~\ref{fig:SecondHarm}, a PNMS experiment is conducted at nearly the avoided crossing resonance condition $V_{dc}\simeq -20$, thus bringing both modal frequencies $\Omega_i$ to the nearest possible value. The figure shows the fundamental PNMS/ATS of both hybrid modes at $\Omega_p=\delta\Omega$.
In addition, we observe a splitting at $\delta\Omega/2$ that can not be explained by the original ATS model~\cite{Autler1955}. This observation suggests the presence of a second parametric coupling tone with the double frequency $2\Omega_p$. Under the linear assumption, the plausible explanation in the current case is the invalidity of the approximation $V_{ac}^2\propto \cos(2\Omega_p t)\approx0$ in the expansion of the forcing term $(V_{ac}+V_{dc})^2$ (see Sec.~S3 and S7~\cite{supp}). Although $V_{ac}^2$ remains by at least one order of magnitude smaller than $2V_{dc}V_{ac}$, its small contribution opens up a splitting at $\Omega_p=\delta\Omega/2$ as seen in the inset of Fig.~\ref{fig:SecondHarm}. Constructing the Floquet Hamiltonian with two-tone drive $\Omega_p,2\Omega_p$ captures both fundamental and second-subharmonic splittings. As shown in Fig.~\ref{fig:SecondHarm}, the theory shows a very good agreement at both the fundamental and the second-subharmonic ATS. Since the model quantitatively captures the second-subharmonic ATS in the linear case, this can provide a useful tool to differentiate between double-frequency parametric excitations, which are explained by the linear model, and nonlinear second-harmonic generation, e.g. at higher amplitudes, in similar systems, see Sec.~S2~\cite{supp} for details.

\textit{Conclusions - }
There have been several attempts to establish dynamical correspondences between quantum and classical systems with time-dependent Hamiltonians~\cite{Lewis1967,Lewis1968,dittrich1994classical,Briggs2013,spreeuw1990classical}. A particularly relevant subclass consists of driven two-level systems with a single-tone parametric coupling, which exhibit the dynamic Stark effect and the associated Autler–Townes splitting at resonance~\cite{Autler1955}. We show that such systems can be mapped onto the slow-manifold dynamics of parametrically coupled classical harmonic oscillators. As a result, under the ATS resonance condition, a splitting of the normal mode takes place in classical systems, which has been studied extensively in nanoelectromechanical and cavity optomechanical systems under the name of parametric normal mode splitting (PNMS). One manifestation of this correspondence is the generation of PNMS at the difference frequency $\delta \Omega$ and its odd fractions. Splittings at $\delta\Omega/2$ are theoretically not permitted for a single-tone parametric drive. We show via Floquet analysis that using a two-tone parametric drive $\Omega_p,2\Omega_p$ in the ATS/PNMS model, however, induces a second-subharmonic ATS at exactly $\delta\Omega/2$. This establishes a connection between second-subharmonic splittings previously observed in parametrically driven nanomechanical systems~\cite{okamoto_coherent_2013,mathew_dynamical_2016} and the generalized ATS framework for bichromatic driving fields or two-photon processes~\cite{Rudolph:98,bakos1977ac}.

In the near-resonant domain $\Omega_p\sim \delta\Omega$, we develop a rigorous model for a system of two parametrically excited coupled harmonic oscillators that realizes PNMS for each of the two oscillators. Our model establishes a direct relation between the width of the normal mode splitting $2g$ and the coupling strength $\lambda$, an aspect which has not been appreciated to date. This provides a tool to quantify the coupling between two modes irrespective of its strength. For the case of weak coupling, this quantity is inaccessible to standard dynamical characterization methods due to the hybridization of the lab coordinates. For strong coupling, an avoided crossing directly reveals the coupling strength if the eigenfrequencies are widely tunable. The presented method overcomes this hurdle. It can 
not only be applied in strain-coupled nanomechanical systems \cite{zalalutdinov_two-dimensional_2006,karabalin_nonlinear_2009,hatanaka_phonon_2014,cha_electrical_2018,stassi_large-scale_2019,doster_observing_2022}, but also in optical or microwave cavity arrays \cite{zhang_synchronization_2012,anderson_engineering_2016}, and in cavity quantum electrodynamics \cite{reithmaier_strong_2004,yoshie_vacuum_2004}, just to name a few. In addition, it can be applied to hybrid systems combining several degrees of freedom, such as spin-mechanical systems~\cite{arcizet_single_2011,lee2017topical}. Here, the presented method supplies a means to detect the coupling to inaccessible dark states, since PNMS can be measured at only one of the two excited modes. 

The theoretical model is complemented by an experimental study using a nanoelectromechanical two-mode system. We study the two orthogonal fundamental flexural modes of a dielectrically controlled SiN nanostring resonator. Dielectric driving is mediated by a dc bias voltage that couples and tunes the eigenmodes. We apply an additional parametric modulation for two different experiments. In the first, we tune the parametric frequency at the difference frequency to induce a PNMS from which the coupling strength is extracted quantitatively in a range of dc bias voltage. In the second experiment, we adjust the bias voltage to the value at which the avoided crossing takes place, then we sweep the parametric drive from 20 to 60 kHz to capture both fundamental and second-subharmonic ATS/PNMS for both hybridized modes. In both cases, we compare the observed splittings with the respective theoretical model and find an excellent traceability of the PNMS. 

All experimental data is available at~\cite{data}. We gratefully acknowledge financial support from the Deutsche Forschungsgemeinschaft
(DFG, German Research Foundation) through Project-ID No. 425217212-SFB 1432 and under 
Germany’s Excellence Strategy—EXC-2111—390814868.
\nocite{griffiths_introduction_2017,Novotny2010,nayfeh_nonlinear_1995,Nayfeh2000,le_3d_2023,bartsch2024}

\bibliography{Biblio_CavOM}

\begin{thebibliography}{66}
\bibitem [{dat()}]{data}%
  \BibitemOpen
  \href@noop {} {}\bibinfo {note} {Experimental data is available at
  10.5281/zenodo.xxxxxxx}\BibitemShut {NoStop}%
\end{thebibliography}

\end{document}


\makeatletter
\renewcommand \thesection{S\@arabic\c@section}
\renewcommand\thetable{S\@arabic\c@table}
\renewcommand \thefigure{S\@arabic\c@figure}
\renewcommand \theequation{S\@arabic\c@equation}
\makeatother


\title{Supplemental Material for \\
Fundamental and second-subharmonic Autler-Townes splitting in classical systems}
\author{Ahmed A. Barakat}
 \email{Contact author: ahmed.barakat@tum.de}
 \affiliation{%
 School of Computation, Information and Technology, Technical University of Munich, Germany.}
 \affiliation{Design and Production Engineering Dept., Faculty of Engineering, Ain Shams University, Egypt.}
\affiliation{Munich Center for Quantum Science and Technology (MCQST), 80799 Munich, Germany}
\affiliation{TUM Center for Quantum Engineering (ZQE), 85748 Garching, Germany}


\author{Avishek Chowdhury}
\affiliation{%
 School of Computation, Information and Technology, Technical University of Munich, Germany.}
 \author{Anh Tuan Le}%
 \affiliation{%
 School of Computation, Information and Technology, Technical University of Munich, Germany.}
\author{Eva M. Weig}
\affiliation{%
 School of Computation, Information and Technology, Technical University of Munich, Germany.}
 \affiliation{Munich Center for Quantum Science and Technology (MCQST), 80799 Munich, Germany}
\affiliation{TUM Center for Quantum Engineering (ZQE), 85748 Garching, Germany}


\date{\today}

\maketitle

\section{Mapping between oscillating-field driving and parametric excitation}

We study the mapping between a classical system of two parametrically coupled harmonic oscillators and the temporal evolution of a quantum two-level system driven by a fast oscillating field. Focusing on the effect of the parametric coupling, a minimal classical system is represented as
\begin{equation}
\ddot q_1+\Omega_1^2 q_1+\Lambda\cos(\Omega_p t)\,q_2=0,\qquad
\ddot q_2+\Omega_2^2 q_2+\Lambda\cos(\Omega_p t)\,q_1=0,
\label{eq:cl_sys}
\end{equation}
where $q_i(t)\in\mathbb{R}$ are the oscillator coordinates, $\Omega_i$ the uncoupled eigenfrequencies, $\Lambda$ the parametric coupling amplitude, and $\Omega_p$ the pump frequency.

The quantum system is described by
\begin{equation}
i\hbar \dot\Psi(t)=[H_0+H_p\cos(\Omega_p t)]\Psi(t),
\label{eq:q_sys}
\end{equation}
with time-independent $H_0$ and $H_p$. We restrict to a two-level subspace $\{|1\rangle,|2\rangle\}$ with
\(
H_0|i\rangle=\hbar\Omega_i|i\rangle
\)
and
\(
\langle i|H_p|j\rangle = 2\hbar\beta
\)
for $i\neq j$, $\langle i|H_p|i\rangle=0$. Writing
\begin{equation}\label{eq:wave_fn}
    \Psi(t)=C_1(t)|1\rangle+C_2(t)|2\rangle,
\end{equation}
and projecting \eqref{eq:q_sys} onto $\langle 1|$ and $\langle 2|$ gives the amplitude equations
\begin{equation}
i\dot C_1(t)=\Omega_1 C_1(t)+2\beta\cos(\Omega_p t)\,C_2(t),\qquad
i\dot C_2(t)=\Omega_2 C_2(t)+2\beta\cos(\Omega_p t)\,C_1(t).
\label{eq:q_amp_sys}
\end{equation}

We now derive the system of the form \eqref{eq:q_amp_sys} from \eqref{eq:cl_sys}. Introduce complex envelopes $A_i(t)$ by
\begin{equation}
q_i(t)=\frac{1}{\sqrt{2\Omega_i}}\Big(A_i(t)e^{-i\Omega_i t}+A_i^*(t)e^{i\Omega_i t}\Big).
\label{eq:qA}
\end{equation}
For each fixed $t$, \eqref{eq:qA} together with its time derivative defines a real–linear invertible change of variables between $(q_i(t),\dot q_i(t))$ and $(A_i(t),A_i^*(t))$.

Substituting \eqref{eq:qA} into \eqref{eq:cl_sys} and collecting the $e^{\mp i\Omega_i t}$ harmonics gives
\begin{equation}
\frac{1}{\sqrt{2\Omega_i}}\Big[(\ddot A_i-2i\Omega_i\dot A_i)e^{-i\Omega_i t}+(\ddot A_i^*+2i\Omega_i\dot A_i^*)e^{i\Omega_i t}\Big]
=
\frac{1}{2\sqrt{2\Omega_j}}\Big(
A_j\big[e^{-i(\Omega_j-\Omega_p)t}+e^{-i(\Omega_j+\Omega_p)t}\big]+A_j^*\big[e^{i(\Omega_j-\Omega_p)t}+e^{i(\Omega_j+\Omega_p)t}\big]\Big),
\label{eq:freepart}
\end{equation}
with $j\neq i$.

We now impose the slow-envelope approximation
\begin{equation}
|\ddot A_i(t)|\ll \Omega_i\,|\dot A_i(t)|,
\end{equation}
which defines the reduced slow dynamics that will be compared to the quantum amplitudes, and project \eqref{eq:freepart} onto the $e^{-i\Omega_i t}$ harmonic, this yields
\begin{equation}\label{eq:Adot}
\begin{aligned}
    -i\dot A_1
    +\frac{\Lambda}{4\sqrt{\Omega_1\Omega_2}}\Big[
      A_2 e^{-i(\Omega_2-\Omega_1+\Omega_p)t}
     +A_2 e^{-i(\Omega_2-\Omega_1-\Omega_p)t}\Big]&=0,\\[2pt]
    -i\dot A_2
    +\frac{\Lambda}{4\sqrt{\Omega_1\Omega_2}}\Big[
      A_1 e^{-i(\Omega_2-\Omega_1+\Omega_p)t}
     +A_1 e^{-i(\Omega_2-\Omega_1-\Omega_p)t}\Big]&=0.
\end{aligned}
\end{equation}
Notice that both terms $\pm\Omega_p$ are present, that is, no rotating wave approximation (RWA) of the Rabi-type is performed. This is especially important for inducing the Bloch-Siegert shift, as described in Section~\ref {sec_perturb}.

Next, define
\begin{equation}
C_i(t)=A_i(t)e^{-i\Omega_i t}.
\end{equation}
Then $A_i(t)=C_i(t)e^{i\Omega_i t}$ and
\begin{equation}
\dot A_i(t)=e^{i\Omega_i t}\big(\dot C_i(t)+i\Omega_i C_i(t)\big).
\label{eq:AdotCidot}
\end{equation}
 Substituting \eqref{eq:AdotCidot} and $A_j(t)=C_j(t)e^{i\Omega_j t}$ into \eqref{eq:Adot}, and cancelling the common exponential factors, gives
\begin{equation}\label{eq:C_Lambda}
    \begin{aligned}
     i\dot C_1(t) &=\Omega_1 C_1(t)+\frac{\Lambda}{2\sqrt{\Omega_1\Omega_2}}\cos(\Omega_p t)\,C_2(t),\\
     i\dot C_2(t) &=\Omega_2 C_2(t)+\frac{\Lambda}{2\sqrt{\Omega_1\Omega_2}}\cos(\Omega_p t)\,C_1(t).
    \end{aligned}
\end{equation}
Within the slow-envelope approximation, the map
\[
(q_i(t),\dot q_i(t))\longleftrightarrow (C_i(t),C_i^*(t))
\]
is therefore a time-dependent linear bijection to the order of the approximation, so the reduced slow dynamics can be expressed equivalently in the complex amplitudes governed by \eqref{eq:C_Lambda}.

Comparing \eqref{eq:q_amp_sys} with \eqref{eq:C_Lambda} gives
\begin{equation}\label{eq:beta-Lambda}
    \beta=\frac{\Lambda}{4\sqrt{\Omega_1\Omega_2}}.
\end{equation}
For this parameter identification, the amplitude equations of the driven two-level system and the slow-amplitude equations of the classical parametric oscillator system coincide, establishing a one-to-one correspondence between the reduced slow dynamics of the classical system and the two-level quantum dynamics.

\section{Floquet Hamiltonian}
The quasi-energies(-frequencies) of either the parametrically excited classical system or the field-perturbed two-level system can be calculated equivalently using Floquet analysis. We begin with the classical system \eqref{eq:cl_sys}. Introduce the ansatz
\begin{equation}\label{eq:Floquet_ansatz_cl}
    q_i(t)=e^{\mu t}\sum_{n}{q_i^{(n)}e^{in\Omega_pt}}, 
\end{equation}
and by spectral decomposition, this yields
\begin{equation}\label{eq:class_quasifreq}
    \Big[\Omega_i^2+(\mu+in\Omega_p)^2\Big]q_i^{(n)}+\frac{\Lambda}{2}\Big[q_j^{(n-1)}+q_j^{(n+1)}\Big]=0.
\end{equation}

This corresponds equivalently to the quantum system \eqref{eq:q_amp_sys}, using the ansatz $C_i=e^{i\lambda t}\sum_n c_i^{(n)} e^{-in\Omega_p t}$, which yields~\cite{Autler1955}
\begin{equation}\label{eq:quant_quasifreq}
    \Big[\lambda +\Omega_i-n\Omega_p\Big]c_i^{(n)}+\beta \Big[c_j^{(n-1)}-c_j^{(n+1)}\Big]=0.
\end{equation}
\begin{figure}[!t]
	\centering
	\includegraphics[width=\linewidth]{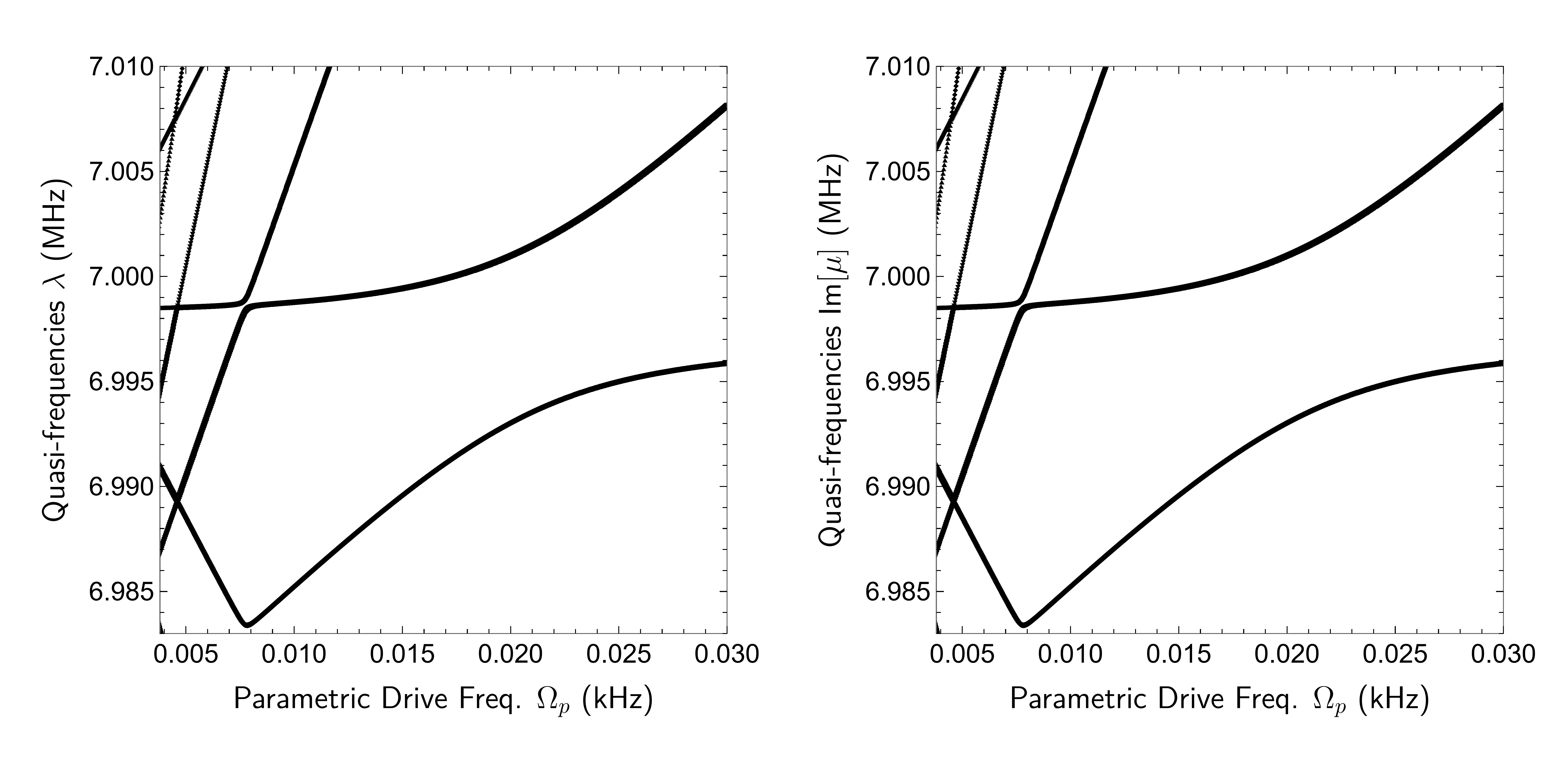}	
	\caption{Quasi-energies showing identical ATS/PNMS structures of the semi-classical quantum system (left, see (\ref{eq:quant_quasifreq})) and the classical system (right, see (\ref{eq:class_quasifreq})), which are identical to the original plot of Autler and Townes~\cite{Autler1955}. }
	\label{fig:Quantclass_floquet}
\end{figure}

In each of the classical or the quantum versions of the problem, this eigenvalue problem is represented by a series in $(n)$ Fourier orders. However, considering the order $n$ of a degree of freedom $i$, the series shows only possible couplings with the directly neighboring $n\pm 1$ of the other degree of freedom $j$. This means that the odd series $n=1,3,5,...$ of $i$ is coupled to the even series $n=2,4,6,..$ of $j$ and vice versa. Nevertheless, the odd and even series of $i$ (and also of $j$) are separable, creating two independent subsystems. They combine to form a \emph{Floquet Hamiltonian} $H_F$~\cite{Shirley_ATS}, in which these subsystems are not coupled by relevant off-diagonal terms. The eigenvalue problem $ H_F\,\boldsymbol{c}=\lambda\,\boldsymbol{c}$ of \eqref{eq:quant_quasifreq} can be represented in matrix form as
\begin{equation}
   \sum_{j=1}^{2}\sum_{m\in \mathbb{Z}}H_{F(i,n),(j,m)}c^{(m)}_j=\lambda c_i^{(n)}, \qquad i\in\{1,2\}.
\end{equation}

The quasi-energies ($\lambda$ or Im[$\mu$]) of both (semi-classical) quantum and classical systems are depicted in Fig.~\ref{fig:Quantclass_floquet}. The ATS/PNMS at $\Omega_p=|\Omega_2-\Omega_1|/n,\,n=1,3,5,...$ show identical structures under the transformation \eqref{eq:beta-Lambda}. However, no splittings can take place at even fractions $\Omega_p=|\Omega_2-\Omega_1|/n,\,n=2,4,...$-- a property well known in the theory~\cite{Autler1955}. This is based on the separation of the even and odd series, as explained earlier. This condition is, however, violated if $H_F$ includes coupling terms between both subsystems. This means that $c_i^{(n)}$ (or $q_i^{(n)}$) are not only related to $c_i^{(n\pm1)}$ but also to $c_i^{(n\pm2)}$, only then, \emph{second-subharmonic ATS/PNMS} can take place as shown in the main text. In this case, the eigenvalue problems for both cases read
    \begin{subequations}\label{eq:quant_quasifreq_secharm}
    \begin{align}
    \Big[\lambda +\Omega_i-n\Omega_p\Big]c_i^{(n)}+\beta \Big[c_j^{(n-1)}-c_j^{(n+1)}\Big]+\beta_{II} \Big[c_j^{(n-2)}-c_j^{(n+2)}\Big]&=0\\
    \Big[\Omega_i^2+(\mu+in\Omega_p)^2\Big]q_i^{(n)}+\frac{\Lambda}{2}\Big[q_j^{(n-1)}+q_j^{(n+1)}\Big]+\frac{\Lambda_{II}}{2} \Big[q_j^{(n-2)}-q_j^{(n+2)}\Big]&=0.
\end{align}
\end{subequations}

An experimental realization of Fig.~\ref{fig:Quantclass_floquet} is shown in the main text for one nanomechanical string resonator, and a plot for another string is found in section~\ref{sec:60muString}, where the splittings are more pronounced but also include nonlinear effects. The physical ground for the appearance of second-subharmonic ATS in our experiment is related to the presence of $|V_{ac}^2|$ as discussed in the main text. In that case the ratio $\Lambda_{II}/\Lambda=|V_{ac}^2|/(2|V_{dc}||V_{ac}|)=|V_{ac}|/(2|V_{dc}|)$ is dictated by the electric field (see~\eqref{Seq:eforce_V}) in the linear case. If the second-subharmonic splitting shows a significantly larger ratio, as in the case presented in Section~\ref{sec:60muString}, this suggests the presence of a nonlinear second-harmonic source. In fact, the spectra of Fig.~\ref{fig:Floquet_Str60} are obtained for relatively higher vibration amplitudes, which supports our inference.

\section{String-System modeling}\label{sec:sys_model}

The dielectrically controlled nanostring represents an electromechanical system that exhibits a strong interaction between the mechanics and the electric field. The nanostring is a continuous dynamical system that has an infinite number of modes in the three dimensional space. However, the most pronounced modes are those corresponding to the in-plane (IP) and out-of-plane (OOP) flexural vibrations, that is the transverse vibrations of the string either in the plane of the chip or the one normal to it, respectively. In each vibration direction, the string exhibits a multitude of modes starting with the fundamental mode, which are dictated by the geometry, material, and pre-stress value. In a previous work, it was observed that the electrostatic field induces a coupling between the fundamental IP and OOP modes~\cite{faust_nonadiabatic_2012}. Starting with the uncoupled two-mode system, it can be described by the equations
\begin{equation}\label{Seq:noelec}
	\begin{aligned}
		\ddot x_1+2\beta \dot x_1+\omega_1^2 x_1=0,\\
		\ddot x_2+2\beta \dot x_2+\omega_2^2 x_2=0,
	\end{aligned}
\end{equation}
where the equations are normalized with respect to the effective mass $m$, $\beta$ is the (normalized) amplitude decay rate, assumed to be similar in both directions, and $\omega_i,\> i=1,2$, are the undamped natural frequencies in both directions. Choosing the IP and OOP directions to represent the modal coordinates, in the case of no surrounding electric field, is evident since these match the principal axes of inertia. Turning to the real problem under study, the nanostring is flanked by two electrodes that are connected to a direct voltage and a smaller alternating voltage. Both electrodes create an electric field $E$ as shown in Fig.~1 of the main text, similar to that of a dipole~\cite{griffiths_introduction_2017}, and the system (\ref{Seq:noelec}) under the electric force $\boldsymbol{F}=[F_1\> F_2]^T$ becomes
\begin{equation}\label{Seq:electromech_def}
	\begin{aligned}
		\ddot x_1+2\beta \dot x_1+\omega_1^2 x_1=F_1,\\
		\ddot x_2+2\beta \dot x_2+\omega_2^2 x_2=F_2,\\
		\boldsymbol{F}=\boldsymbol{p}.\nabla\boldsymbol{E}=\alpha \nabla \boldsymbol{E}^2,
	\end{aligned}
\end{equation}
where $\boldsymbol{p}=\alpha \boldsymbol{E}$ is the dipole moment created by dielectric polarization in the silicon nitride string due to the applied electric field $\boldsymbol{E}$, whereas $F$ is the force acted upon by the electric field. Moreover, the result of the last equation in (\ref{Seq:electromech_def}) relies on the fact that $\nabla \times \boldsymbol{E}=0$ up to a negligible time-varying magnetic field, yielding $\boldsymbol{E}.\nabla \boldsymbol{E}=\nabla \boldsymbol{E}^2$.\\

The positioning of the nanostring, at its equilibrium position, off-centered between the electrodes as shown in Fig.~1 of the main text, plays an important role in the system dynamics. Upon application of a voltage, this off-centered position exposes the nanostring to an inhomogeneous electric field, which allows it to generate restoring forces since $\partial_{x_i} F_j \neq 0,\> i,j\in\{1,2\}$. For this reason, the electric force in each direction $i$ is expanded as
\begin{equation}\label{Seq:eforce_exp}
	F_i=F_{0i}+\frac{\partial F_i}{\partial x_i}x_i+\frac{\partial F_i}{\partial x_j}x_j,\quad i\neq j.
\end{equation}

Elaborating on the excitation signals, the potential difference $V$ across the electrodes that induces the electric field is composed of direct and alternating parts $V_{dc}, V_{ac}$, respectively. However, for $V_{ac}\ll V_{dc}$, then 
\begin{equation}\label{Seq:eforce_V}
	F_i(\boldsymbol{E}^2)=c_iV^2\simeq c_{i}(V_{dc}^2+2V_{dc}V_{ac}).
\end{equation}

Inserting both (\ref{Seq:eforce_exp}) and (\ref{Seq:eforce_V}) in (\ref{Seq:electromech_def}) yields
\begin{equation}\label{Seq:electromech_exp}
\begin{aligned}
	\ddot x_1 + 2\beta \dot x_1+ \omega_1^2 x_1 -(c_{11}V_{dc}^2+2c_{11}V_{dc}V_{ac}(t))x_1-(c_{12}V_{dc}^2+2c_{12}V_{dc}V_{ac}(t))x_2=&c_{01}V_{dc}^2+2c_{01}V_{dc}V_{ac}(t),\\
	\ddot x_2 + 2\beta \dot x_2+ \omega_1^2 x_2 -(c_{21}V_{dc}^2+2c_{21}V_{dc}V_{ac}(t))x_1-(c_{22}V_{dc}^2+2c_{22}V_{dc}V_{ac}(t))x_2=&c_{02}V_{dc}^2+2c_{02}V_{dc}V_{ac}(t),
\end{aligned}
\end{equation}
where deducing the coefficients $c_{ij}$ from field calculations is left for a separate work. From a dynamical point of view, each term in this expansion plays a different role: in the left hand side, $c_{ii}V_{dc}^2$ act as restoring forces, while  $c_{ij}V_{dc}^2$ represent the linear coupling forces, in addition to producing  intramodal and intermodal parametric excitation terms $2c_{ii}V_{dc}V_{ac}(t),2c_{ij}V_{dc}V_{ac}(t)$, respectively. Furthermore, the right-hand side terms describe the static and dynamical forcing drive. From (\ref{Seq:electromech_exp}), and originally from (\ref{Seq:eforce_V}), it can be obviously noticed that any time-varying excitation leads inherently not only to a forced drive but also to a parametric excitation. Nevertheless, as ascertained experimentally, the effect of the parametric excitation on the system's response is hardly observed unless the excitation frequency is tuned to any of the parametric resonance frequencies. Thereby, the parametric excitation can be safely considered as a perturbation of the stiffness terms.\\

The system (\ref{Seq:electromech_exp}) can be made more readable by considering some simplifications: removing the static forcing terms $c_{0i}V_{dc}^2$ since they do not contribute to the system dynamics; the dynamic forcing terms are also represented as $f(t)$ for both coordinates, since their amplitudes do not change the system dynamics, only their same excitation frequency; $c_{21}$ and $c_{12}$ are assumed equal relying on having a conservative symmetry-preserving electric field. Moreover, the terms can be further reduced, yielding

\begin{equation}\label{Seq:electromech_red}
	\begin{aligned}
		\ddot x_1 + 2\beta \dot x_1+ (\omega_1^2+\Delta\omega_1^2) x_1 +\eta_1(t)x_1+(\lambda^2+\eta_c(t))x_2=&f(t),\\
		\ddot x_2 + 2\beta \dot x_2+ (\omega_2^2+\Delta\omega_2^2) x_2 +\eta_2(t)x_2+ (\lambda^2+\eta_c(t))x_1=&f(t),
	\end{aligned}
\end{equation}
where $\Delta\omega_i^2=-c_{ii}V_{dc}^2,\>\eta_i(t)=-2c_{ii}V_{dc}V_{ac}(t),\>\lambda^2=c_{\lambda}V_{dc}^2$ and $\eta_c(t)=2c_{\lambda}V_{dc}V_{ac}(t)$, where $-c_{ij}$ is rewritten as $c_\lambda$ for brevity.

In this view, (\ref{Seq:electromech_red}) represents a two-mode, i.e. two-degree-of-freedom, dynamical system with intra- and intermodal parametric excitations, $\eta_i(t)$ and $\eta_c(t)$, respectively, in addition to a linear coupling $\lambda$.

Here, it is essential to distinguish between system terms and excitation terms to determine the system's eigenmodes. Thus, by excluding the excitation terms, we regard the autonomous version of the system to be
\begin{equation}\label{Seq:electromech_auto}
	\begin{aligned}
		\ddot x_1 + 2\beta \dot x_1+ \tilde\omega_1^2 x_1+\lambda^2 x_2=&0,\\
		\ddot x_2 + 2\beta \dot x_2+ \tilde\omega_2^2 x_2 +\lambda^2 x_1=&0,
	\end{aligned}
\end{equation}
where $\tilde{\omega}_i^2=\omega_i^2+\Delta\omega_i^2$. 

In addition to the mathematical convenience, the following experimental investigation requires transforming the system from the current coordinates into its modal coordinates, which is influenced by the presence of the electrically induced coupling $\lambda$. To this end, we use the matrix of eigenvectors 
\begin{equation}
    U=\begin{bmatrix}
        \frac{1}{2}(\tilde{\omega}_1^2-\tilde{\omega}_2^2)-\frac{1}{2}\sqrt{(\tilde{\omega}_2^2-\tilde{\omega}_1^2)^2 + 4\lambda^4}&&\lambda^2\\
        \frac{1}{2}(\tilde{\omega}_1^2-\tilde{\omega}_2^2)+\frac{1}{2}\sqrt{(\tilde{\omega}_2^2-\tilde{\omega}_1^2)^2 + 4\lambda^4}&&\lambda^2
    \end{bmatrix}
\end{equation}
where $v_i$ are the eigenvectors of (\ref{Seq:electromech_auto}), in which $\boldsymbol{q}=U\boldsymbol{x}$, yielding
\begin{equation}\label{Seq:electromech_hybrid}
	\begin{aligned}
		\ddot q_1 + 2\gamma \dot q_1+ \Omega_1^2 q_1=&0,\\
		\ddot q_2 + 2\gamma \dot q_2+ \Omega_2^2 q_2=&0,
	\end{aligned}
\end{equation}
where $\gamma$ is obtained empirically, and
\begin{equation}\label{Seq:eigfreq_hybrid}
	\Omega^2_{1,2}=\frac{1}{2}(\tilde{\omega}_1^2+\tilde{\omega}_2^2)\mp\frac{1}{2} \sqrt{(\tilde{\omega}_2^2-\tilde{\omega}_1^2)^2 + 4\lambda^4}.
\end{equation}
Applying the transformation $U$ to the excitation terms in (\ref{Seq:electromech_red}) yields the whole system in the modal coordinates as follows
\begin{equation}\label{Seq:electromech_assym}
	\begin{aligned}
		\ddot q_1 + 2\gamma \dot q_1+ (\Omega_1^2+\Sigma_1(t)) q_1+\zeta\Lambda(t) q_2=&F(t),\\
		\ddot q_2 + 2\gamma \dot q_2+ (\Omega_2^2+\Sigma_2(t)) q_2+\>\>\Lambda(t) q_1=&F(t),
	\end{aligned}
\end{equation}
where
\begin{equation*}\label{Seq:param_exc_coeff}
\begin{aligned}
	\Sigma_1&= \left(\frac{1}{2} + \frac{\delta\tilde{\omega}^2}{2\sqrt{\delta\tilde{\omega}^4+ 4 \lambda^4}}\right) \eta_1+ \left(\frac{1}{2} - \frac{\delta\tilde{\omega}^2}{
    2 \sqrt{\delta\tilde{\omega}^4 + 4 \lambda^4}}\right) \eta_2-\left(\frac{2 \lambda^2}{\sqrt{\delta\tilde{\omega}^4 + 4 \lambda^4}}\right)\eta_c,\\
    \Sigma_2&= \left(\frac{1}{2} - \frac{\delta\tilde{\omega}^2}{2\sqrt{\delta\tilde{\omega}^4+ 4 \lambda^4}}\right) \eta_1+ \left(\frac{1}{2} + \frac{\delta\tilde{\omega}^2}{
    2 \sqrt{\delta\tilde{\omega}^4 + 4 \lambda^4}}\right) \eta_2+\left(\frac{2 \lambda^2}{\sqrt{\delta\tilde{\omega}^4 + 4 \lambda^4}}\right)\eta_c,\\
    \Lambda &=\left(-\frac{1}{2}+\frac{\delta\tilde{\omega}^2}{2\sqrt{\delta\tilde{\omega}^4+4\lambda^4}}\right) \eta_1+\left(\frac{1}{2}-\frac{\delta\tilde{\omega}^2}{2\sqrt{\delta\tilde{\omega}^4+4\lambda^4}}\right) \eta_2+ \frac{\delta\tilde{\omega}^2}{2 \lambda^2} \left(-1+\frac{\delta\tilde{\omega}^2}{\sqrt{\delta\tilde{\omega}^4+4\lambda^4}}\right)\eta_c,
    \end{aligned}
\end{equation*}
with a scaling factor $\zeta=\frac{1}{2 \lambda^4}(\delta\tilde{\omega}^4+2\lambda^4+\delta\tilde{\omega}^2\sqrt{\delta\tilde{\omega}^4+4\lambda^4})$, where $\delta\tilde{\omega}^2=\tilde{\omega}_2^2-\tilde{\omega}_1^2$ and $\delta\tilde{\omega}^4=(\delta\tilde{\omega}^2)^2$ for brevity. Thus, (\ref{Seq:electromech_assym}) represents a system of two damped oscillators with the intramodal excitation terms $\Sigma_1,\Sigma_2$ and a symmetric coupling, \emph{up to a scaling factor $\zeta$}, via the parametric coupling terms $\Lambda$.

Since the estimation of the coupling strength is the end goal of this work, as will be shown later, a direct transformation between $\Lambda$ and the coupling strength $|\lambda^2|=c_{\lambda}V_{dc}^2$ is then essential, \emph{since the parametric normal modes splitting (PNMS) vanishes if $\Lambda$ does}. Thereby, unfolding the reduced symbols used in (\ref{Seq:electromech_red}) transforms $\Lambda$ in (\ref{Seq:electromech_assym}) into
\begin{equation}\label{Seq:Lambda_lambda_rel}
	\Lambda=\frac{V_{ac}(t)\delta\omega^2}{V_{dc}}\left(\frac{\delta\omega^2+(c_{11}-c_{22})V_{dc}^2}{\sqrt{4\lambda^4+((c_{11}-c_{22})V_{dc}^2+\delta\omega^2)^2}}-1\right),
\end{equation}
where $\delta\omega^2=\omega_2^2-\omega_1^2$. From this relation it can be then ascertained that the presence of the parametric coupling $\Lambda$ inside the modal coordinates relies primarily on the presence of a coupling between the unexcited mechanical modes $\lambda$, since  if $\lambda=0$ then $\Lambda=0$, in addition to the time-varying excitation $V_{ac}$ and the degeneracy between the natural frequencies $\omega_1,\omega_2$. In other words, the vanishing of $c_{11}$ or $c_{22}$ does not eliminate $\Lambda$, and thereby the whole phenomenon. This proves that the modal coupling is \emph{a necessary condition} for the PNMS, and thereby the PNMS can be directly attributed to the presence of the modal coupling in the physical coordinates $x_1,x_2$. The expression of $\Lambda$ in (\ref{Seq:electromech_assym}) shows that the phenomenon can take place for any of the parametric excitation mechanisms in the physical coordinates, namely $\eta_1(t),\eta_2(t)$ or $\eta_c(t)$. In this regard, the PNMS discussed in Okamoto et al.~\cite{okamoto_coherent_2013} presents a special case, where only the intramodal excitation $\eta_1(t)$ exists. However, in case of existing $c_{ii}$ terms, as in our experiments, their values can determined by an additional experiment involving the frequency tuning of the system's eigenvalues, since the coefficients $c_{ii}$ determine both the intramodal parametric excitation terms $\eta_1,\eta_2$ as well as the frequency tuning terms $\Delta \omega_1, \Delta \omega_2$. We provide an example of their estimation for our physical system using a parabolic function confined to a limited range around $V_{dc}=0$ in Fig.~3 of the main text.

\section{Relation between PNMS and the avoided level crossing}
As outlined in the introduction of the main article, both the PNMS and the avoided level crossing refer to the coupling between the mechanical vibration modes of the resonator. The PNMS requires an additional parametric pump, but it is not limited to the point of resonance ($\tilde{\omega}_1=\tilde{\omega}_2$). However, the avoided crossing occurs only at this resonance condition and does not require any additional parametric excitation. Thereby, a relation should exist between both phenomena at resonance. This could be explained by considering the avoided crossing opening, using (\ref{Seq:eigfreq_hybrid}),
\begin{equation}
  \Gamma= |\Omega_2-\Omega_1|=\sqrt{\frac{1}{2}(\tilde{\omega}_1^2+\tilde{\omega}_2^2)+\frac{1}{2} \sqrt{(\tilde{\omega}_2^2-\tilde{\omega}_1^2)^2 + 4\lambda^4}}- \sqrt{\frac{1}{2}(\tilde{\omega}_1^2+\tilde{\omega}_2^2)-\frac{1}{2} \sqrt{(\tilde{\omega}_2^2-\tilde{\omega}_1^2)^2 + 4\lambda^4}},
\end{equation}
and for $\tilde{\omega}_1=\tilde{\omega}_2=\tilde{\omega}$,
\begin{equation}
    \Gamma^2=2\tilde{\omega}^2-2\sqrt{\tilde{\omega}^4-\lambda^4},
\end{equation}
for $\lambda^4 \ll \tilde{\omega}^4$
\begin{equation}\label{Seq:AC-CStr_approx}
    \Gamma\approx \frac{\lambda^2}{\tilde{\omega}},
\end{equation}
which matches, under the preceding condition, the definition in Refs.~\cite{Novotny2010,faust_nonadiabatic_2012}. However, it should be noted that the inclined spring model used in these works does not correspond to the general case studied in this work, where the stiffness tuning $\Delta \omega_i^2=-c_{ii}V_{dc}^2$ does not equal the coupling rate $\lambda^2=c_{\lambda}V_{dc}^2$. Notice that we have used the notation $\Gamma$ as in \cite{Novotny2010,faust_nonadiabatic_2012} rather than $2g$ (cf. the introduction of the main article) for the opening of the avoiding crossing, which serves in our case to clearly differentiate between this phenomenon and the PNMS. 

Since the coupling is manifested in two different phenomena, the condition of strong coupling will be revisited. As discussed in the introduction and the literature cited therein, the coupling (here $\lambda$) between two modes is named strong when it exceeds the dissipation rate. However, in relation to the avoided crossing $\Gamma$, it should be larger than the line-width of the coupled modes $2\gamma$. For the given approximation in~(\ref{Seq:AC-CStr_approx}) on resonance, the condition for strong coupling then yields
\begin{equation}
    \lambda\gg \sqrt{2\gamma \tilde{\omega}}.
\end{equation}
\section{Dynamical perturbation analysis}\label{sec_perturb}
To analyze the dynamics of a generic parametrically excited coupled two-mode system  in its modal coordinates, (\ref{Seq:electromech_assym}) will be the starting point, only while assuming $\zeta=1$, which will be shown to only have a scaling effect on the results. The dynamics are studied using the multiple-scales perturbation method~\cite{nayfeh_nonlinear_1995}. By perturbing the system using a small parameter $\epsilon \ll 1$, this reads
\begin{equation}\label{Seq:electromech_symm}
	\begin{aligned}
		\ddot q_1 + 2\epsilon\>\gamma \dot q_1+ (\Omega_1^2+\epsilon\>\Sigma_1(t)) q_1+\epsilon\>\Lambda(t) q_2=&\epsilon \>F(t),\\
		\ddot q_2 + 2\epsilon\>\gamma \dot q_2+ (\Omega_2^2+\epsilon\>\Sigma_2(t)) q_2+\epsilon\>\Lambda(t) q_1=&\epsilon \>F(t),
	\end{aligned}
\end{equation}
with $\Sigma_i(t)=\Sigma_{i}\cos(\Omega_p t),\Lambda(t)=\Lambda\cos(\Omega_p t)$ and $F(t)=F_0\cos(\Omega_f t)$, ignoring phase differences.

One seeks an expansion up to the second order in the form

\begin{equation}\label{Seq:perturbed_sol}
	\begin{aligned}
		q_1(t;\epsilon)=q_{10}(T_0,T_1,T_2)+\epsilon q_{11}(T_0,T_1,T_2)+\epsilon^2 q_{12}(T_0,T_1,T_2),\\
		q_2(t;\epsilon)=q_{20}(T_0,T_1,T_2)+\epsilon q_{21}(T_0,T_1,T_2)+\epsilon^2 q_{22}(T_0,T_1,T_2),
	\end{aligned}
\end{equation}

where $T_i=\epsilon^i t$ are the different time scales, whereas the differential operator for each time scale is defined as $D_i=\partial/\partial T_i$, this means
\begin{equation}\label{Seq:perturbed_diff}
	\begin{aligned}
		\frac{\partial}{\partial t}&=D_0+\epsilon D_1+\epsilon^2D_2,\\
		\frac{\partial^2}{\partial t^2}&=D_0^2+2\epsilon D_0 D_1+\epsilon^2(D_1^2+D_0D_2).
	\end{aligned}
\end{equation} 

Inserting~(\ref{Seq:perturbed_sol}) and~(\ref{Seq:perturbed_diff}) in (\ref{Seq:electromech_symm}) produces a set of equations which can be separated with respect to the perturbation order $\epsilon^i$. For ${\epsilon^0}$, the unperturbed system reads
\begin{equation}\label{Seq:Ord_Null}
	\begin{aligned}
		D_0^2q_{10}+\Omega_1^2 q_{10}=0,\\
		D_0^2q_{20}+\Omega_2^2 q_{20}=0,
	\end{aligned}
\end{equation}

while for a perturbation up to the first order, i.e. ${\epsilon^1}$, the following equations are deduced
\begin{equation}\label{Seq:Ord_Eins}
	\begin{aligned}
		D_0^2q_{11}+\Omega_1^2 q_{11}=-2\gamma D_0q_{10}-\Sigma_{1} q_{10} \cos(\Omega_p t)-\Lambda q_{20} \cos(\Omega_p t)-2D_0D_1q_{10}+F_0\cos(\Omega_f t),\\
			D_0^2 q_{21}+\Omega_2^2 q_{21}=-2 \gamma D_0 q_{20}-\Lambda q_{10} \cos(\Omega_p t)-\Sigma_{2} q_{20} \cos(\Omega_p t)-2D_0D_1 q_{20}+F_0\cos(\Omega_f t),
	\end{aligned}
\end{equation}

and finally for the second order ${\epsilon^2}$, we get
\begin{equation}\label{Seq:Ord_Zwei}
	\begin{aligned}
		\begin{split}
			D_0^2q_{12}+\Omega_1^2 q_{12}&=-2\gamma (D_0q_{11}+D_1q_{10})-\Sigma_{1} q_{11} \cos(\Omega_p t)-\Lambda q_{21} \cos(\Omega_p t)\\&-D_1^2q_{10}-2D_0D_2q_{10}-2D_0D_1q_{11},
		\end{split}\\
		\begin{split}
			D_0^2q_{22}+\Omega_2^2 q_{22}&=-2\gamma (D_0q_{21}+D_1q_{20})-\Sigma_{2} q_{21} \cos(\Omega_p t)-\Lambda q_{11} \cos(\Omega_p t)\\&-D_1^2q_{20}-2D_0D_2q_{20}-2D_0D_1q_{21}.
		\end{split}
	\end{aligned}
\end{equation}

The solution of~(\ref{Seq:Ord_Null}) can be written as
\begin{equation}\label{Seq:Null_Loes}
	\begin{aligned}
		q_{10}(T_0,T_1,T_2)=A_1(T_1,T_2)e^{i\Omega_1T_0}+\bar{A_1}(T_1,T_2)e^{-i\Omega_1T_0},\\
		q_{20}(T_0,T_1,T_2)=A_2(T_1,T_2)e^{i\Omega_2T_0}+\bar{A_2}(T_1,T_2)e^{-i\Omega_2T_0},
	\end{aligned}
\end{equation}
where $A_1,A_2$ represent the slowly varying amplitudes, with slow time-scales $T_1,T_2$, whereas the exponential functions represent the fast oscillations.\\

Like any perturbation method, the perturbation solutions are sought locally around a steady-state solution (\ref{Seq:Null_Loes}), which can be determined exactly. Since the perturbation expansion series (\ref{Seq:perturbed_sol}) is assumed to be asymptotic, the terms appearing in (\ref{Seq:Ord_Eins}) can not cause instability in the higher order perturbation terms $q_{11},q_{21},...$. For this reason, resonating terms in these equations are called \emph{secular terms}, which are actually functions of the $A_1,A_2$, and are accordingly equated to zero. This adds a constraint on the slowly varying amplitudes, thereby dictating amplitude-frequency relationships~\cite{Nayfeh2000}.\\

To apply this to the problem at hand, the solution (\ref{Seq:Null_Loes}) is inserted in (\ref{Seq:Ord_Eins}) to solve for the perturbation dynamics of the first order. Moreover, the excitation frequencies are detuned around the resonance frequencies, such that
\begin{equation}\label{Seq:detun}
	\Omega_f= \Omega_1+\epsilon \sigma_f,\quad \Omega_p= \Omega_2-\Omega_1+\epsilon\sigma_p,
\end{equation}
to produce the steady state solution, by detuning $\Omega_f$, and to evoke the PNMS through detuning $\Omega_p$ around the difference frequency. The right-hand side in $A_1,A_2$ should not include secular terms to avoid divergence of the perturbation solutions $q_{11},q_{21}$. Thus, the secular terms are extracted and equated to zero, that is
\begin{equation}\label{Seq:solcond_ord1}
	\begin{aligned}
		2i\Omega_1D_1A_1+2i\gamma\Omega_1A_1+\frac{\Lambda}{2}e^{-i\sigma_pT_1}A_2&=\frac{F_0}{2}e^{i\sigma_f T_1},\\
		2i\Omega_2D_1A_1+2i\gamma\Omega_1A_2+\frac{\Lambda}{2}e^{-i\sigma_pT_1}A_1&=0.
	\end{aligned}
\end{equation}

Moreover, by returning to (\ref{Seq:Ord_Eins}) and eliminating the secular terms, solutions can be found for $q_{11}$ and $q_{21}$, which is irrelevant to this discussion. Nevertheless, using these solutions in addition to those of (\ref{Seq:Null_Loes}) can be afterwards inserted in (\ref{Seq:Ord_Zwei}) to repeat the same procedure for the second order. The secular terms are again extracted and equated to zero forming the \emph{solvability conditions} for the second-order equations, which read
\begin{equation}\label{Seq:solcond_ord2}
	\begin{aligned}
D_1^2A_1+2\gamma D_1A_1+2i\Omega_1D_2A_1+\frac{\Lambda^2}{16\Omega_1^2-16\Omega_1\Omega_2}A_1-\frac{\Sigma_{1}^2}{6\Omega_1^2+4\Omega_1\Omega_2-\Omega_2^2}A_1&=0,\\
D_1^2A_2+2\gamma D_1A_2+2i\Omega_2D_2A_2+\frac{\Lambda^2}{16\Omega_2^2-16\Omega_1\Omega_2}A_2-\frac{\Sigma_{2}^2}{6\Omega_1^2+4\Omega_1\Omega_2-\Omega_2^2}A_2&=0,
	\end{aligned}
\end{equation}
which seem to be decoupled, but due to (\ref{Seq:solcond_ord1}) they are actually not.\\

The slow time-scale dynamics are now described by (\ref*{Seq:solcond_ord1}) and (\ref{Seq:solcond_ord2}) but in different orders of approximation. Combining them together by expanding the total time derivative of the slowly varying amplitudes yields
\begin{equation}\label{Seq:totderiv_Amp}
	\begin{aligned}
		\frac{dA_1}{dt}&=\epsilon D_1 A_1+\epsilon^2 D_2 A_1\\
		&=\left(-\epsilon\gamma-\frac{i\epsilon^2}{32\Omega_1^2}\left[16\gamma^2\Omega_1+\frac{8\Sigma_{1}^2\Omega_1}{(3\Omega_1-\Omega_2)(\Omega_1+\Omega_2)}+\Lambda^2(\frac{1}{\Omega_2}+\frac{1}{\Omega_2-\Omega_1})\right]\right)A_1\\
		&+i\left(\frac{\epsilon }{4\Omega_1}+\frac{\epsilon^2\sigma_p}{8\Omega_2^2}\right)\Lambda e^{-i\epsilon\sigma_p t}A_2+\frac{F_0}{8\Omega_1^2}\left(\epsilon^2\gamma+i(\epsilon^2\sigma_f-2\epsilon\Omega_1)\right)e^{i\epsilon\sigma_ft}\\
		\frac{dA_2}{dt}&=\epsilon D_1 A_2+\epsilon^2 D_2 A_2\\
		&=\left(-\epsilon\gamma-\frac{i\epsilon^2}{32\Omega_2^2}\left[16\gamma^2\Omega_2-\frac{8\Sigma_{2}^2\Omega_2}{(3\Omega_2-\Omega_1)(\Omega_1+\Omega_2)}+\Lambda^2(\frac{1}{\Omega_1}+\frac{1}{\Omega_1-\Omega_2})\right]\right)A_2\\
		&+i\left(\frac{\epsilon }{4\Omega_2}+\frac{\epsilon^2\sigma_p}{8\Omega_1^2}\right)\Lambda e^{-i\epsilon\sigma_p t}A_1+\frac{\epsilon^2 \Lambda F_0}{32\Omega_1\Omega_2^2}e^{i\epsilon(\sigma_p+\sigma_f)t}
	\end{aligned}
\end{equation}
The dynamics of the slowly varying amplitudes in (\ref{Seq:totderiv_Amp}) include time-varying system parameters; however, they can be better understood in a rotating frame according to
\begin{equation}\label{Seq:rotframe}
	\begin{aligned}
		A_1(t)&=\frac{1}{2}a_1e^{-i\epsilon\sigma_p t/2},\\
		A_2(t)&=\frac{1}{2}a_2e^{i\epsilon\sigma_p t/2},
	\end{aligned}
\end{equation}
where they then read
\begin{equation}\label{Seq:rotframe_Amp}
	\begin{aligned}
	a'_1&=\frac{\epsilon}{32}\left(-32\gamma-\frac{16i\epsilon\gamma^2}{\Omega_1}+i\epsilon\Omega_1^2\left[16\sigma_p\Omega_1^2-\frac{8\Sigma_{1}^2\Omega_1}{(3\Omega_1-\Omega_2)(\Omega_1+\Omega_2)}-\Lambda^2(\frac{1}{\Omega_2}+\frac{1}{\Omega_1-\Omega_2})\right]\right)a_1\\
		&+i\frac{\epsilon(\epsilon\sigma_p+2\Omega_1)}{8\Omega_1^2}\Lambda a_2+\frac{\epsilon F_0}{4\Omega_1^2}\left(\epsilon\gamma+i(\epsilon\sigma_f-2\Omega_1)\right)e^{i\epsilon(\sigma_p+2\sigma_ft)/2},\\
		a'_2&=\frac{\epsilon}{32}\left(-32\gamma-\frac{16i\epsilon\gamma^2}{\Omega_2}-i\epsilon\Omega_2^2\left[16\sigma_p\Omega_2^2-\frac{8\Sigma_{2}^2\Omega_2}{(3\Omega_1-\Omega_2)(\Omega_1+\Omega_2)}+\Lambda^2(\frac{1}{\Omega_1}+\frac{1}{\Omega_1-\Omega_2})\right]\right)a_2\\
		&-i\frac{\epsilon(-\epsilon\sigma_p+2\Omega_2)}{8\Omega_1^2}\Lambda a_1+\frac{\epsilon^2 \Lambda F_0}{32\Omega_1\Omega_2^2}e^{i\epsilon(\sigma_p+2\sigma_f)t/2}.
	\end{aligned}
\end{equation}

\begin{figure}
		\includegraphics[width=0.8\textwidth]{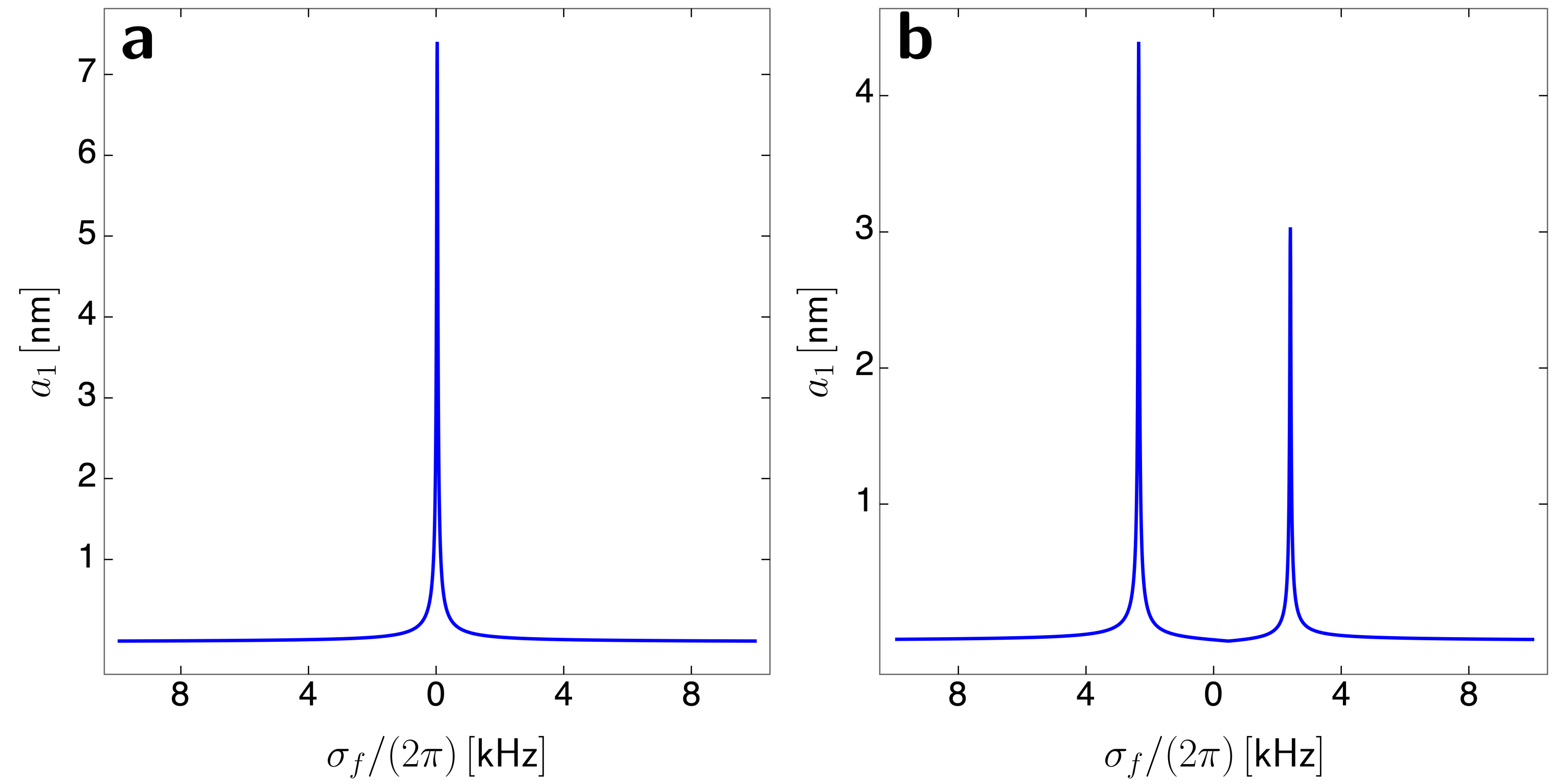}
	\caption{Simulation of the parametric normal mode splitting for $\Omega_1/(2\pi)=6.71\,\text{MHz},\Omega_2/(2\pi)=6.75\,\text{MHz},\Sigma_{1}=\Sigma_{2}=0,F_1=10\,\text{N/kg},\gamma/(2\pi)=100\,\text{Hz}$, where (a)  no parametric drive is applied, (b) a parametric drive is tuned at $\sigma_p=\Omega_p-(\Omega_2-\Omega_1)=0$ with  an amplitude of $\Lambda/(2\pi)^2=10^{10}\text{Hz}^2$. The former depicts the unperturbed resonance of mode $1$, whereas the latter reveals the PNMS.} 
	\label{Sfig:Sim_PNMS}
\end{figure}

Equations (\ref{Seq:rotframe_Amp}) represent, therefore, the slow dynamics occurring when the PNMS is evoked. They constitute a two-mode dynamical system with external drive. This means that the solution of such a system would be a combination of transient and steady-state particular solutions. The steady-state solution of $a_1$, which is too long to be stated, is then calculated and presented in Fig.~\ref{Sfig:Sim_PNMS} in the absence and presence of a parametric excitation, respectively.\\

For the actuated nanostring case in (\ref{Seq:electromech_exp}), it must be emphasized that the parametric excitation takes place inherently due to any time-varying electrical field. Thereby, toggling between the existence and non-existence in Fig.~\ref{Sfig:Sim_PNMS} can not mean inducing a forced excitation without a parametric excitation; however, the existence of this phenomenon depends upon choosing the correct parametric excitation frequency. Accordingly, the phenomenon can be evoked when the parametric excitation is adjusted to the difference frequency, i.e. $\Omega_p\simeq|\Omega_2-\Omega_1|$. Thereby, Fig.~\ref{Sfig:Sim_PNMS}(right) shows the effect of inducing the phenomenon through a correct tuning of the parametric excitation frequency. \\

\begin{figure}
\centering
		\includegraphics[width=0.8\textwidth]{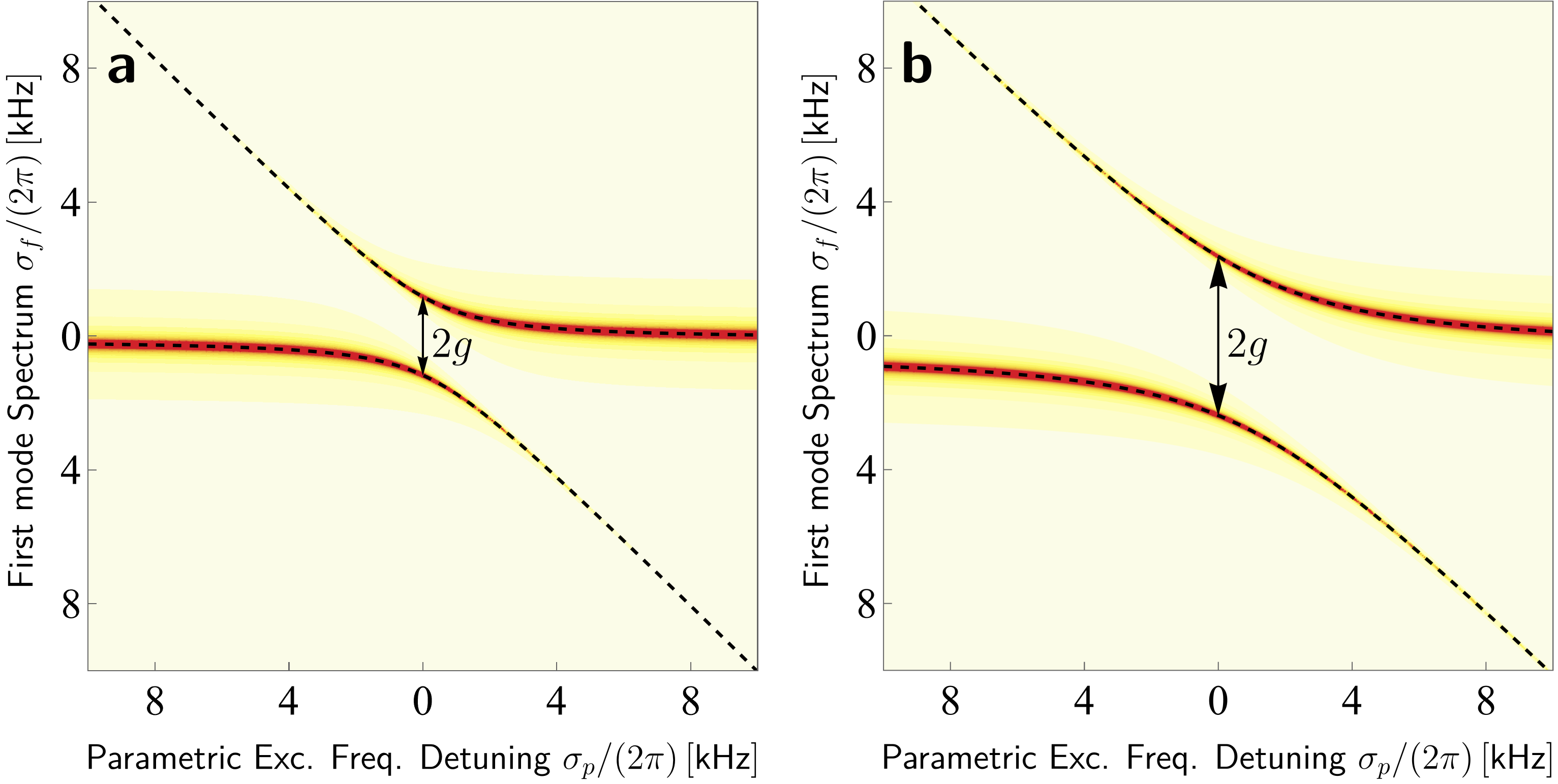}
	\caption{Calculated PNMS showing the spectrum of the first mode with amplitude $a_1$ on a color scale as a function of the detuning of the parametric drive from the difference frequency $\Omega_2-\Omega_1$ for two different values of the parametric excitation amplitude (a) $\Lambda/(2\pi)^2=0.5\times10^{10}\text{Hz}^2$, (b) $\Lambda/(2\pi)^2=10^{10}\text{Hz}^2$. The change of the width of the splitting $2g$ is clearly apparent. The dashed lines represent the loci of the first mode's splitted branches according to Eq.~(\ref{Seq:splitlocus}).
 }
	\label{Sfig:PNMS_Lambda}
\end{figure}

A more holistic picture is obtained for each value of $\sigma_p$ by viewing the PNMS inside the parameter space $\sigma_p-\sigma_f$, where the amplitudes are represented in a color scale in Fig.~\ref{Sfig:PNMS_Lambda}. In this figure, the effect of the parametric excitation amplitude $\Lambda$ on the splitting width is also presented. On the other hand, it was found that the intramodal excitation amplitudes $\Sigma_{1},\Sigma_{2}$ do not have any effect on the splitting width. Since the splitting width is the measurable quantity through the experiments, the focus of this theoretical work will be given to the proper use of the parameter $\Lambda$ for the sake of estimating the coupling strength. This direct relation between $\Lambda$ and the coupling strength $\lambda$ was deduced in (\ref{Seq:Lambda_lambda_rel}).\\

 The splitting width $2g$ is defined here to be the difference between the two branches of the modal resonance frequency $\Omega_f=\Omega_1+\sigma_f$ at exactly the parametric resonance $\sigma_p=0$. The splitting width is quantified by extracting the locus of each of the two resonance frequency branches. The two loci are deduced through tracing the local maxima of the two branches. By simplifying the expressions for the local maxima and assuming $\Sigma_{1}=\Sigma_{2}=\gamma=0$ for their minimal effects, we find
\begin{equation}\label{Seq:splitlocus}
\begin{split}
	\sigma_{f\pm}&=\frac{1}{32\Omega_1^2\Omega_2^2\delta\Omega}\Biggl(-(\delta\Omega(16\sigma_p \Omega_1^2\Omega_2^2+\Lambda^2(\Omega_1+\Omega_2))\\ &\pm\sqrt{256\sigma_p^2\Omega_1^4\Omega_2^4\delta\Omega^2+\Lambda^4(\Omega_1^2-\Omega_1\Omega_2+\Omega_2^2)^2+16\Lambda^2\Omega_1^2\Omega_2^2\delta\Omega(2\sigma_p\Omega_1\Omega_2+4\Omega_1\Omega_2\delta\Omega-\sigma_p^2\delta\Omega))}\Biggr),
\end{split}
\end{equation}
where $\delta\Omega=\Omega_1-\Omega_2$. This expression is plotted additionally in Fig.~\ref{Sfig:PNMS_Lambda} as dashed lines, and despite the simplification it shows a very good agreement with the numerical results.\\

Mathematically, the splitting width $2g$ is obtained by setting $\sigma_p=0$ in (\ref{Seq:splitlocus}) and calculating the difference $2g=|\sigma_{f+}-\sigma_{f-}|$ yielding
\begin{equation}\label{Seq:split_width}
	2g=|\frac{\sqrt{\zeta}\Lambda\sqrt{64\Omega_1^3\Omega_2^3\delta\Omega^2+\zeta\Lambda^2(\Omega_1^2-\Omega_1\Omega_2+\Omega_2^2)^2}}{16\Omega_1^2\Omega_2^2\delta\Omega}|,
\end{equation}
 where $\Lambda$ is shown to be scaled by $\sqrt{\zeta}$ as discussed below the Eq. \eqref{Seq:electromech_assym}. Taking the scaling factor $\sqrt{\zeta}$ into consideration, i.e. $\Lambda_\zeta=\sqrt{\zeta}\Lambda$, suggests rewriting \eqref{Seq:split_width} as
\begin{equation}\label{Seq:split_width_scaled}
	2g=|\frac{\Lambda_\zeta\sqrt{64\Omega_1^3\Omega_2^3\delta\Omega^2+\Lambda_\zeta^2(\Omega_1^2-\Omega_1\Omega_2+\Omega_2^2)^2}}{16\Omega_1^2\Omega_2^2\delta\Omega}|.
\end{equation}
 An interesting result is obtained by the reordering of \eqref{Seq:split_width}. By setting 
 \begin{equation}
     \tilde{\Lambda}=\Lambda_\zeta/(2\sqrt{\Omega_1\Omega_2}),
 \end{equation}
one finds
\begin{equation}
    2g=\tilde{\Lambda}\sqrt{1+\frac{\tilde{\Lambda}^2}{16}\frac{(\Omega_1^2-\Omega_1\Omega_2+\Omega_2^2)^2}{\Omega_1^2\Omega_2^2\delta\Omega^2}}.
\end{equation}
Then expand the second term under the root and consider $\delta\Omega\ll \Omega_1,\Omega_2$, this yields
\begin{equation}
    2g\simeq\tilde{\Lambda}\sqrt{1+\Big(\frac{\tilde{\Lambda}}{4\delta\Omega}\Big)^2},
\end{equation}
where $\tilde{\Lambda}$ is the Rabi frequency and $(\tilde{\Lambda}/(4\delta\Omega))^2$ is the Bloch-Siegert shift.

\begin{figure}
\centering
	\includegraphics[width=0.45\textwidth]{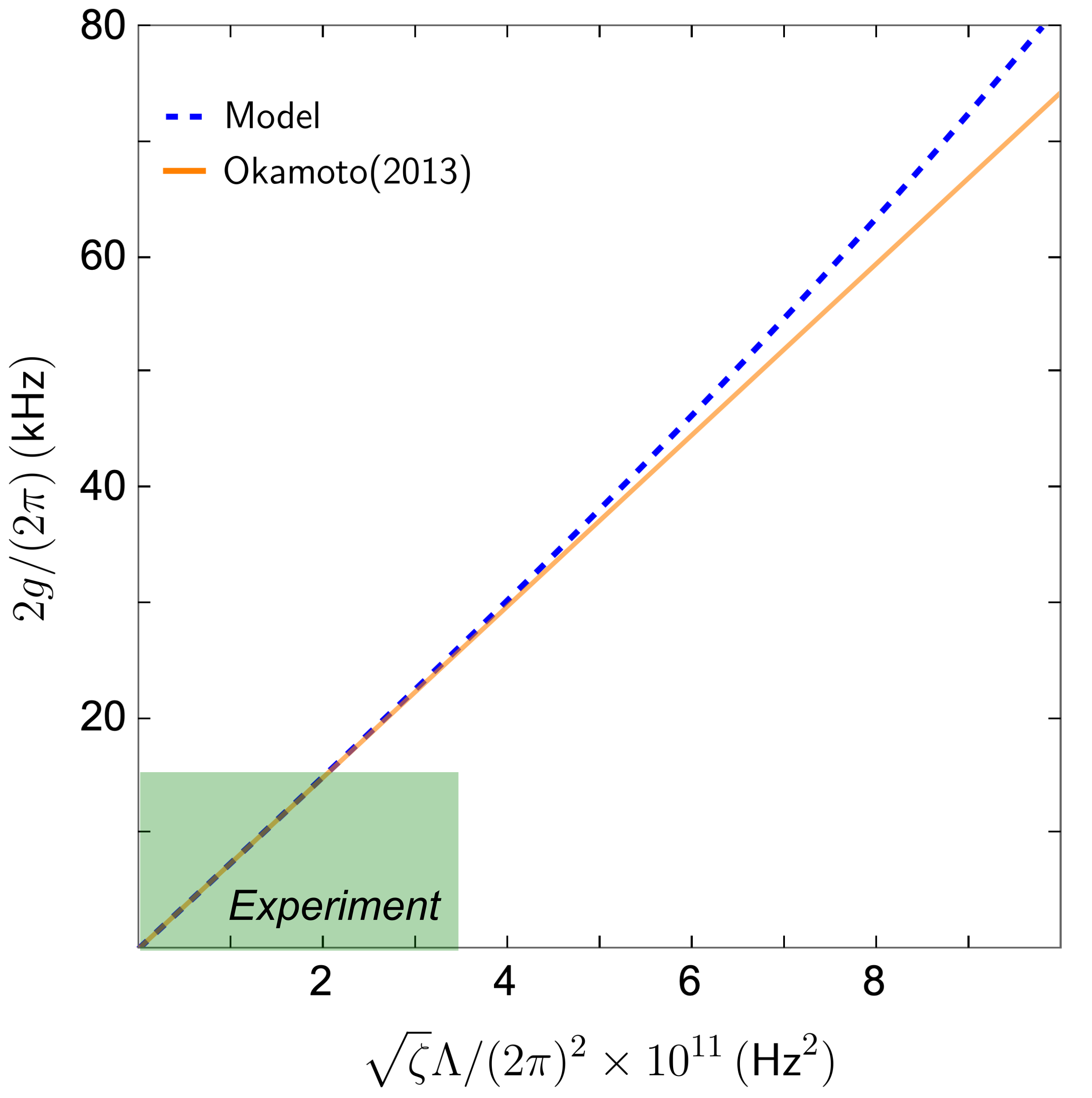}	
	\caption{Width of the PNMS as a function of the parametric drive strength assuming $\Omega_1/(2\pi)=6.71$MHz, $\Omega_2/(2\pi)=6.75$MHz (\ref{Seq:split_width}). The blue dashed line depicts this function according to our PNMS model in Eq.~(\ref{Seq:split_width}), while the orange solid line displays the approximate solution Eq.~(\ref{Seq:split_okamoto}) as discussed in~\cite{okamoto_coherent_2013}. The relevant parameter range of the experiment is indicated by the light green box.}
	\label{Sfig:g_compare}
\end{figure}

A comparable theoretical study was carried out by Okamoto et al~\cite{okamoto_coherent_2013}, however, a full analytical expression for the splitting loci was not provided. Nevertheless, an expression, up to the first approximation, for the splitting width is given to be
\begin{equation}\label{Seq:split_okamoto}
	2g \approx \tilde{\Lambda}=\frac{\Lambda_\zeta}{2\sqrt{\Omega_1\Omega_2}},
\end{equation}
which is readily reproduced by (\ref{Seq:split_width}) by eliminating the second term in the square root. A graphical comparison between (\ref{Seq:split_width}) and the approximate value (\ref{Seq:split_okamoto}) from Ref.~\cite{okamoto_coherent_2013} is presented in Fig.~\ref{Sfig:g_compare}. Clearly, (\ref{Seq:split_okamoto}) represents a good approximation for relatively small parametric drive amplitude $\Lambda$.

In order to extract the coupling strength $\lambda$ from the splitting width $2g$ of the PNMS the loci (\ref{Seq:splitlocus}) are fitted to the experimental data to extract the splitting width $2g$. The parametric coupling $\Lambda$ is calculated according to (\ref{Seq:split_width}). The calculated $\Lambda$ is then inserted in (\ref{Seq:Lambda_lambda_rel}) to determine the coupling strength $\lambda$ with the help of the parameters determined from Fig.~1(b) as discussed in the main text. The calculation is carried out in two steps due to the complexity of each expression. Another calculation method relying on the assumption $\lambda=\sqrt{c_\lambda}V_{dc}$ (see its definition in~\eqref{Seq:electromech_red}), can also be used as explained in the main text, resulting in Fig.~3. That figure shows an excellent agreement between the assumption and the experimental data. The fitting details are explained in Section~\ref{sec:Fitting}. In this case, the $c_\lambda$ constant becomes the fitting parameter and is used afterwards to determine $\lambda$ at any $V_{dc}$. Both methods yield the same results within an acceptable tolerance.

\begin{figure}[!t]
	\centering
	\includegraphics[width=0.5\linewidth]{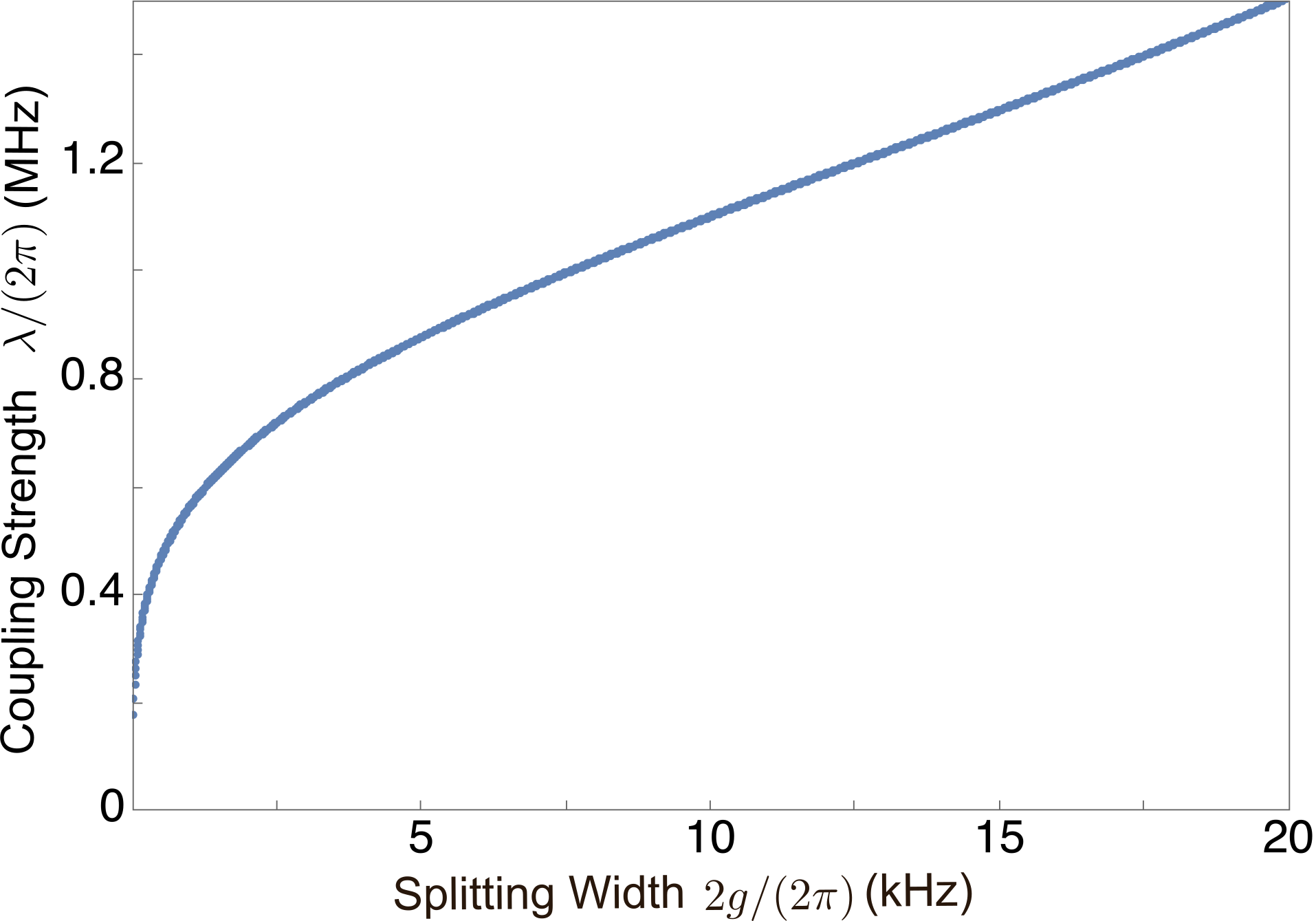}	
	\caption{Illustration of the relation between the system’s
coupling strength $\lambda$ against the splitting width $2g$ according to the approximate expression in \eqref{Seq:g_lambda_simp} for $V_{ac}$ = 1 V, $V_{dc}$ = 10 V, $\omega_1/(2\pi)$ = 6.63 MHz and $\omega_2/(2\pi)$ = 6.85 MHz.}
	\label{fig:Lambda_theo_exp}
\end{figure}

For the sake of illustration, we deduce an approximate direct relation between the splitting width $2g$ and the implicit coupling strength $\lambda$ up to three assumptions. It neglects the intramodal excitation coefficients $c_{ii}$, the scaling factor $\zeta$ and the second-order term in $\Lambda$ in (\ref{Seq:split_width}). Under these assumptions, combining (\ref{Seq:eigfreq_hybrid}),(\ref{Seq:Lambda_lambda_rel}) and (\ref{Seq:split_okamoto}) yields
\begin{equation}\label{Seq:g_lambda_simp}
    2g\approx\frac{-V_{ac}(\omega_2^2-\omega_1^2)}{V_{dc}(\omega_1^2\omega_2^2-\lambda^4)^{1/4}}\left[\frac{\omega_2^2-\omega_1^2}{\sqrt{(\omega_2^2-\omega_1^2)^2+4\lambda^4}}-1\right],
\end{equation}
where $\omega_i$, as defined before, are the fundamental frequencies in the physical coordinates (\ref{Seq:electromech_exp}). Considering the range of splitting widths observed in measurements, the approximate values of the corresponding coupling strength are depicted in Fig.~\ref{fig:Lambda_theo_exp}. The figure shows the direct relation between the splitting width and the coupling strength. This highlights the edge of the applied method in retrieving a rather elusive coupling strength using PNMS through a generic mathematical framework. The significance of this finding relies on the fact that the whole analysis is not limited to the underlying physics, which is applicable to any two-mode system experiencing a linear modal coupling.
\section{Measurement system}
\begin{figure}[!ht]
	\centering
	\includegraphics[width=0.4\textwidth]{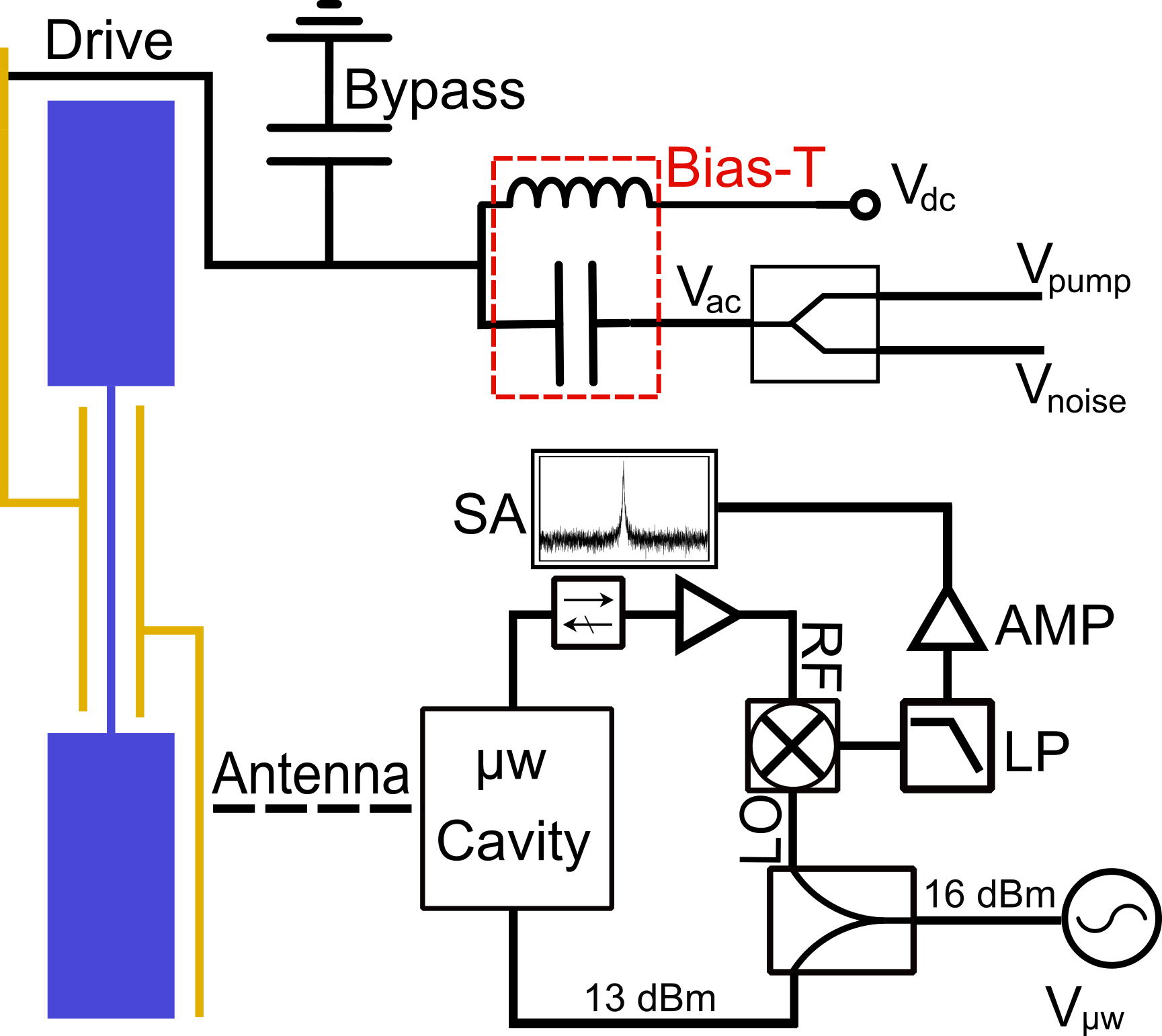}	
	\caption{Schematic of the nanostring along with its transduction circuitry.}
	\label{Sfig:messsystem}
\end{figure}
The $65\,\mu$m strongly pre-stressed SiN string has been fabricated on a fused silica substrate along with two adjacent gold electrodes. A schematic of the dielectric transduction setup is shown in Fig.~\ref{Sfig:messsystem}. The string is actuated dielectrically by applying an rf and a dc voltage $V_{dc}$, which are combined using a bias tee. The rf signal is also a combination of both the parametric pump $V_{\text{pump}}$ and a noise drive $V_{\text{noise}}$, which then together constitute $V_{ac}$. A bypass single layer capacitor is used to ground the microwave signal while allowing the rf and dc signals to pass through for one of the gold electrodes around the string. The other electrode is bonded to an antenna which capacitively couples to one of the modes of a $\lambda/4$ coaxial microwave cavity for measurement enhancement~\cite{le_3d_2023}. A signal generator is tuned on the resonance frequency of the microwave cavity $3.088$\,GHz with output power of $16$\,dBm, providing the cavity with only $13$\,dBm via a power splitter. The mechanical motion, thereby, modulates the cavity's natural frequency, acting as a time-varying capacitor in an analogous LC circuit powered by a resonating signal $V_{\mu w}$. This generates two sidebands in the microwave cavity signal. The signal passes through an rf isolator and amplifier before being fed into an IQ-mixer. Mixing with a reference signal (LO) from the microwave signal generator results in a frequency down-conversion of all spectral components of the signal. All high-frequency mixing products are filtered out using a low-pass filter. The remaining heterodyne signal at the mechanical vibration frequency is amplified before readout using a signal analyzer (SA).

\section{Amplitude-Splitting relation}
\begin{figure}[!ht]
	\centering
	\includegraphics[width=0.5\textwidth]{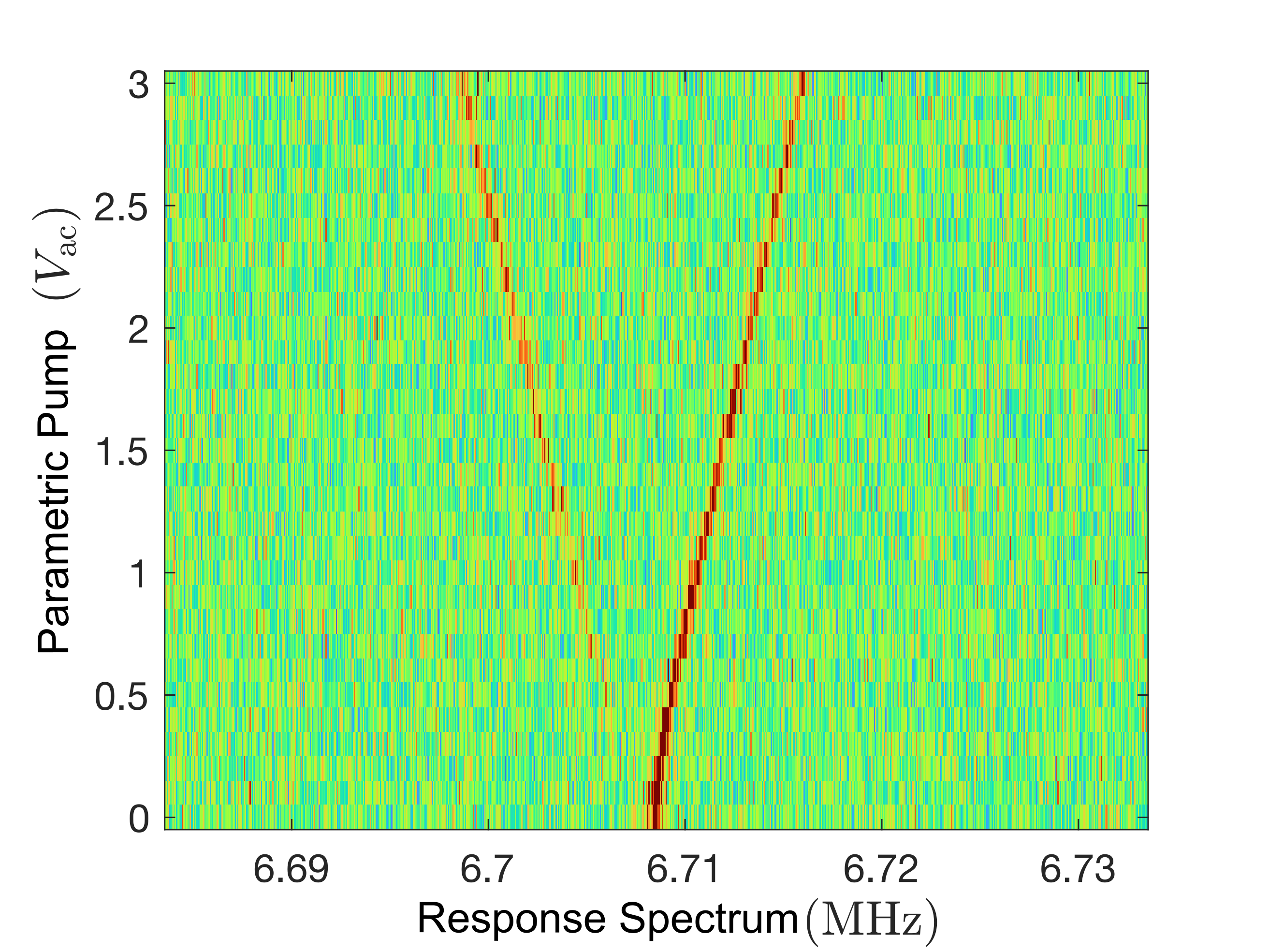}	
	\caption{PNMS of mode $1$ at $V_{dc}=-20$ V as a function of parametric modulation amplitude.}
	\label{Sfig:branching}
\end{figure}

The relation between the excitation amplitude $V_{ac}$ (with a constant $V_{\text{noise}}$) and the response spectrum is shown in Fig.~\ref{Sfig:branching}. The depicted map shows the linear dependence of the splitting width $2g$ on the excitation amplitude as hypothesized in our theory. This linearity, and hence our model, therefore applies to our analysis of the data since the value of the used parametric perturbation is $V_{ac}=2\>V$. Two main conclusions can be drawn from this figure. Firstly, the PNMS splitting is shown to take place as long as the parametric excitation exists; this means that there is no threshold for the phenomenon. Secondly, the relation is shown to be linear in the range of our experiments which implies the linearity of the two relations deduced in the proposed theoretical model: (i) $\Lambda \propto V_{ac}$ in (\ref{Seq:Lambda_lambda_rel}) and (ii) $2g \propto \Lambda$ in (\ref{Seq:split_width}), which renders the second order term in the latter relation negligible within the voltage interval used in this work. Moreover, this also implies the linear approximation in (\ref{Seq:eforce_V}), otherwise nonlinear direct and parametric excitation terms would have propagated in the following analysis up to (\ref{Seq:Lambda_lambda_rel}) as well as in the perturbation analysis. 
\section{Fitting of data in Fig.3}\label{sec:Fitting}
The data in Fig.3 (main text) represent the calculated $\Lambda_\zeta$ values according to~\eqref{Seq:split_width_scaled}. They are then fitted using Eq.4 (main text), which is the $\zeta$-scaled version of~\eqref{Seq:Lambda_lambda_rel}. The fitting assumes a linear relation between $\lambda$ and $V_{dc}$ $\lambda=\sqrt{c_\lambda}V_{dc}$ (see~\eqref{Seq:electromech_red}). This enables finding the coupling tuning parameter $\sqrt{c_\lambda}$. Since $\tilde{\omega}_2-\tilde{\omega}_1=\delta\omega+(c_{11}-c_{22})=0$ at the avoided crossings, the first term of~\eqref{Seq:Lambda_lambda_rel} and Eq.4 then vanishes. This permits the correction of the small inaccuracies in $c_{11}$ and $c_{22}$ independently since they determine $V_{dc}$ of the avoided crossings exactly. Additionally, at these two points, the actual value of the external drive voltage $V_{ac}$ at the terminals of the electrodes can be determined separately.
\section{Experiment on a $60\,\mu$m string}\label{sec:60muString}
\begin{figure}[!ht]
	\centering
	\includegraphics[width=0.5\textwidth]{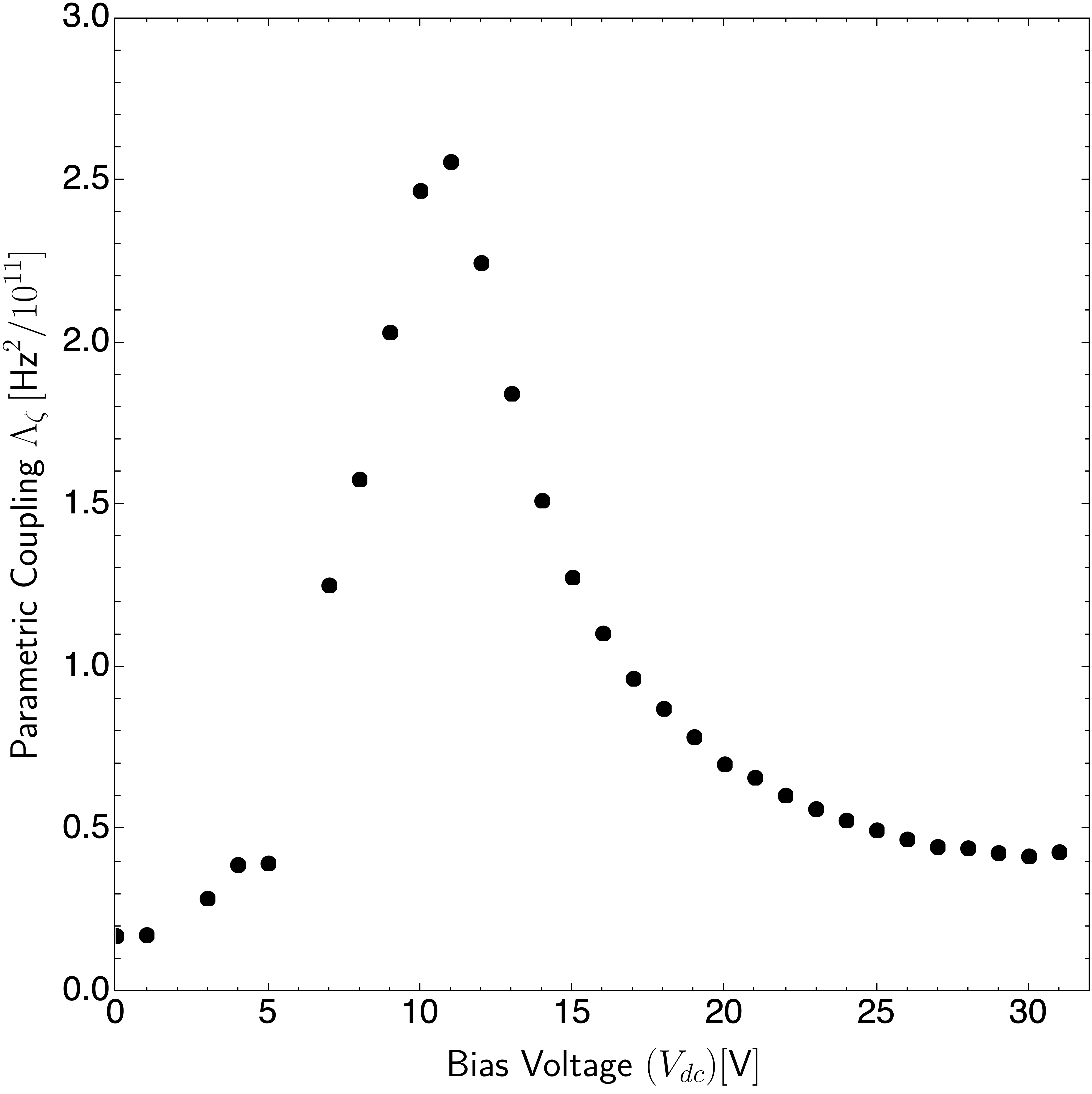}	
	\caption{Parametric coupling $\Lambda_\zeta$ as a function of applied dc bias voltage obtained for the $60\,\mu$m-long string.}
	\label{Sfig:60mustring}
\end{figure}

The calculations and experiments presented in this work for the $65\,\mu$m SiN string are repeated for a shorter $60\,\mu$m one on the same device. The scaled parameter coupling $\Lambda_\zeta$ dependence of the shorter string on the absolute value of the applied  bias voltage is depicted in Fig.~\ref{Sfig:60mustring}, qualitatively confirming the findings presented in Fig.~3 of the main text. 

The $60\,\mu$m string was used in another study to deduce the coupling strength $\lambda$ using a different numerical approach~\cite{bartsch2024}. 
The results reveal that the deduced coupling coefficient from the optimization approach in Ref.~\cite{bartsch2024} yields a mean value of $0.688$\,MHz, whereas using the present approach, a coupling strength of $0.675$\,MHz is calculated out of the corresponding $\Lambda_\zeta$. Given the fact that they are methodologically independent approaches for measuring the same value for the same string using exactly the same setup (except for the excitation signal) and still obtaining an agreement of less than $5$\% allows for a significant mutual support of both approaches.

\begin{figure}[!t]
	\centering
	\includegraphics[width=0.99\linewidth]{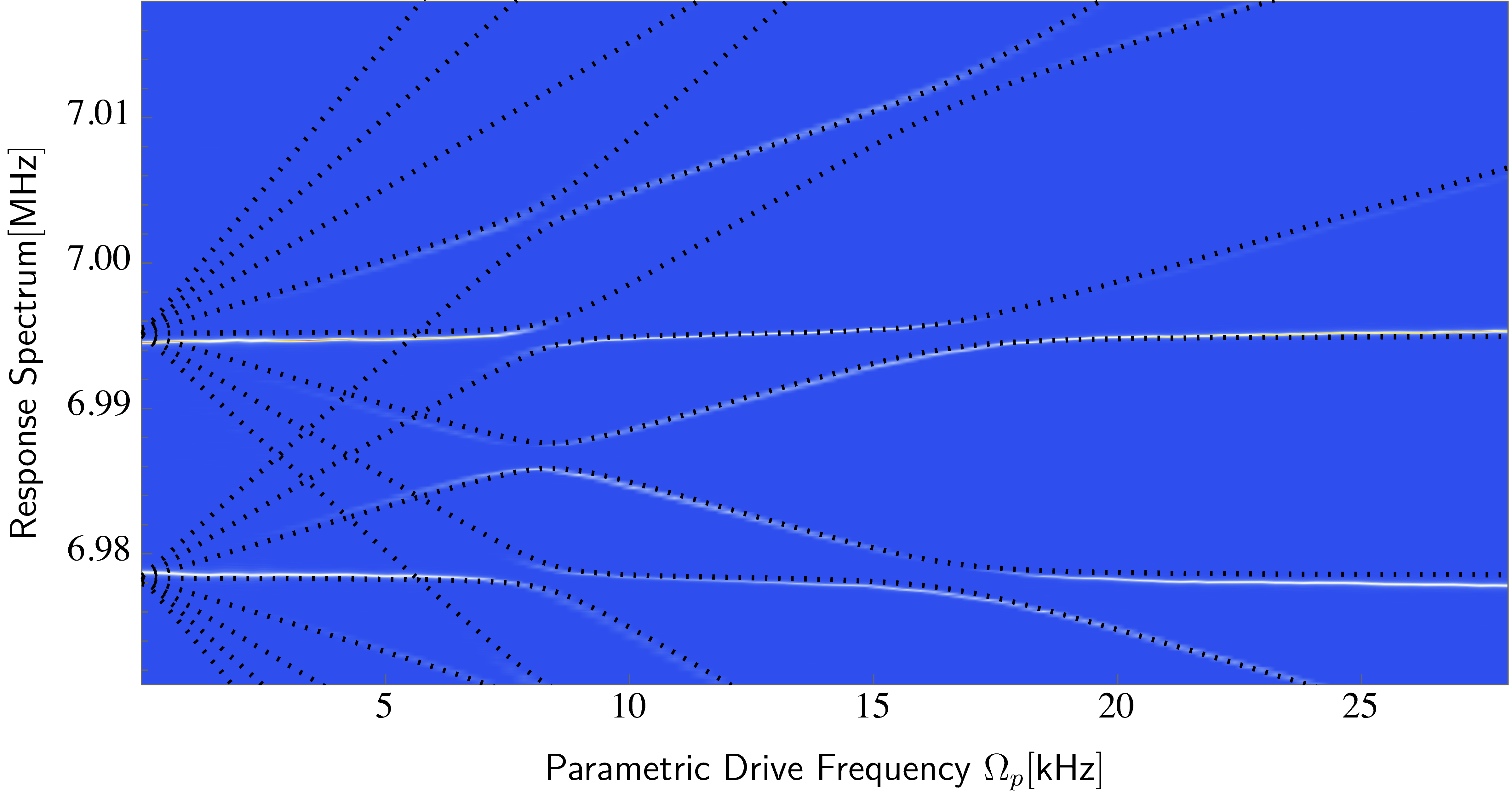}	
	\caption{Experimental map of the quasi-energies at the avoided crossing point $V_{dc}\simeq 18\,V$ by sweeping the parametric frequency, showing the fundamental ATS at $\Omega_p=\delta \Omega\simeq17$kHz and the second-subharmonic ATS at $\delta \Omega/2\simeq8.5$kHz. The dashed black lines show the Floquet model for a two-tone drive. }
	\label{fig:Floquet_Str60}
\end{figure}

Moreover, the Floquet quasienergies are also measured for this string near the avoided crossing point and shown in Fig.~\ref{fig:Floquet_Str60}. Both $\Omega_1$ and $\Omega_2$ modes show fundamental, second-subharmonic and other subharmonic ATS. The ratio of second-subharmonic to fundamental ATS $\Lambda_{II}/\Lambda$ (see~\eqref{eq:quant_quasifreq_secharm}) appears to be significantly larger than that ratio for the $65\mu$m string, which is presented in the main text. This is attributed to the presence of nonlinear second-harmonic generation.

\bibliography{Biblio_CavOM}